\documentclass{article} 
\usepackage{tencent_ailab_tech_report}

\usepackage[T1]{fontenc} 
\usepackage[utf8]{inputenc} 

\usepackage{microtype}
\usepackage{hyperref}
\usepackage{url}
\usepackage{booktabs}
\definecolor{darkblue}{rgb}{0, 0, 0.5}
\hypersetup{colorlinks=true, citecolor=darkblue, linkcolor=darkblue, urlcolor=darkblue}

\usepackage{graphicx}
\usepackage{enumitem}
\usepackage{pgfplots}
\usepackage{tabularx}
\usepackage{caption}
\usepackage{subcaption}
\usepackage{bbm}
\usepackage{url}
\usepackage{amsmath} 
\usepackage{pifont}
\usepackage{arydshln}
\usepackage{makecell}
\usepackage{booktabs}
\usepackage{multirow} 
\usepackage{algorithm}
\usepackage{algorithmic}

\usepackage[most]{tcolorbox}
\usepackage[frozencache,cachedir=.]{minted}

\DeclareUnicodeCharacter{211D}{\ensuremath{\mathbb{R}}}
\DeclareUnicodeCharacter{2124}{\ensuremath{\mathbb{Z}}}
\DeclareUnicodeCharacter{21A6}{\ensuremath{\mapsto}}

\DeclareUnicodeCharacter{2264}{\ensuremath{\leq}}
\DeclareUnicodeCharacter{2265}{\ensuremath{\geq}}
\DeclareUnicodeCharacter{2192}{\ensuremath{\rightarrow}}
\DeclareUnicodeCharacter{2194}{\ensuremath{\leftrightarrow}}
\DeclareUnicodeCharacter{2227}{\ensuremath{\wedge}}
\DeclareUnicodeCharacter{22A2}{\ensuremath{\vdash}}
\DeclareUnicodeCharacter{2080}{\ensuremath{_0}}
\DeclareUnicodeCharacter{2081}{\ensuremath{_1}}
\DeclareUnicodeCharacter{2082}{\ensuremath{_2}}
\DeclareUnicodeCharacter{2083}{\ensuremath{_3}}
\DeclareUnicodeCharacter{2084}{\ensuremath{_4}}
\DeclareUnicodeCharacter{2085}{\ensuremath{_5}}
\DeclareUnicodeCharacter{2086}{\ensuremath{_6}}
\DeclareUnicodeCharacter{2087}{\ensuremath{_7}}
\DeclareUnicodeCharacter{2088}{\ensuremath{_8}}
\DeclareUnicodeCharacter{2089}{\ensuremath{_9}}
\DeclareUnicodeCharacter{2200}{\ensuremath{\forall}}
\DeclareUnicodeCharacter{2203}{\ensuremath{\exists}}
\DeclareUnicodeCharacter{2228}{\ensuremath{\lor}}
\DeclareUnicodeCharacter{2260}{\ensuremath{\neq}}
\DeclareUnicodeCharacter{230B}{\ensuremath{\rfloor}} 
\DeclareUnicodeCharacter{2223}{\ensuremath{\mid}}    
\DeclareUnicodeCharacter{230A}{\ensuremath{\lfloor}} 
\DeclareUnicodeCharacter{2090}{\ensuremath{_a}}  
\DeclareUnicodeCharacter{2091}{\ensuremath{_e}}  
\DeclareUnicodeCharacter{2092}{\ensuremath{_o}}  
\DeclareUnicodeCharacter{2093}{\ensuremath{_x}}  
\DeclareUnicodeCharacter{2211}{\ensuremath{\sum}}        
\DeclareUnicodeCharacter{2115}{\ensuremath{\mathbb{N}}}  

\DeclareUnicodeCharacter{2208}{\ensuremath{\in}} 
\title{Towards Solving More Challenging IMO Problems via Decoupled Reasoning and Proving}

\author{Zhenwen Liang, Linfeng Song, Yang Li, Tao Yang, Feng Zhang, Haitao Mi and Dong Yu \\
Tencent AI Lab\\
\texttt{\{zhenwzliang,lfsong,haitaomi\}@global.tencent.com} \\
}

\colmfinalcopy 
\begin{document}

\maketitle

\begin{abstract}
Automated Theorem Proving (ATP) in formal languages is a foundational challenge for AI. While Large Language Models (LLMs) have driven remarkable progress, a significant gap remains between their powerful informal reasoning capabilities and their weak formal proving performance. Recent studies show that the informal accuracy exceeds 80\% while formal success remains below 8\% on benchmarks like PutnamBench. We argue this gap persists because current state-of-the-art provers, by tightly coupling reasoning and proving, are trained with paradigms that inadvertently punish deep reasoning in favor of shallow, tactic-based strategies.
To bridge this fundamental gap, we propose a novel framework that decouples high-level reasoning from low-level proof generation. Our approach utilizes two distinct, specialized models: a powerful, general-purpose \emph{Reasoner} to generate diverse, strategic subgoal lemmas, and an efficient \emph{Prover} to rigorously verify them. This modular design liberates the model's full reasoning potential and bypasses the pitfalls of end-to-end training. We evaluate our method on a challenging set of post-2000 IMO problems, a problem set on which no prior open-source prover has reported success. Our decoupled framework successfully solves 5 of these problems, demonstrating a significant step towards automated reasoning on exceptionally difficult mathematical challenges. To foster future research, we release our full dataset of generated and verified lemmas for a wide range of IMO problems, available at \url{https://tencent-imo.github.io/}.
\end{abstract}

\section{Introduction}

Automated Theorem Proving (ATP) is the task of automatically generating formal proofs for mathematical or logical statements. By translating problems into a formal language (e.g., Lean \citep{moura2021lean} or Isabelle \citep{paulson1994isabelle}) and iteratively applying tactics within a proof assistant's environment, an ATP system can construct machine-verified proofs that guarantee logical correctness. This verifiability makes ATP indispensable for the formal verification of critical software and hardware systems, where every reasoning step must be rigorously checked. ATP has long been a foundational challenge in both AI and mathematics, as such systems could leverage massive computational power to help mathematicians evaluate and even solve \emph{open conjectures}.
Recent breakthroughs in large language models (LLMs) have catalyzed rapid progress in ATP. Leveraging techniques such as expert iteration \citep{polu2020generative}, tree search \citep{hunyuan_prover, intern25_stepprover,bfs_prover,liang2025mps}, chain-of-thought reasoning \citep{ds_prover_v15, goedel_prover}, and reinforcement learning, state-of-the-art provers have achieved remarkable performance gains. 

However, recent research in the LLM reasoning community has exposed a significant and widening gap between informal reasoning and formal proving capabilities. For instance, a recent evaluation \citep{dekoninck2025open} on the PutnamBench illustrates this chasm: \emph{Top-tier LLMs like Gemini 2.5 Pro can generate informal, natural-language solutions with over 80\% accuracy after human verification. In contrast, the best formal prover Deepseek-Prover-V2 671B struggles to solve even 8\% of the same problems directly.}
This highlights a critical dilemma for the field: while LLMs possess immense mathematical reasoning power, their informal outputs lack the rigor and machine-verifiable guarantees required for formal correctness. Conversely, formal ATP systems have this guarantee but lack the raw problem-solving ability to tackle complex challenges. Bridging this gap is arguably the most pressing challenge in modern ATP.
To address this, state-of-the-art provers like DeepSeek-Prover-v2 \citep{ds_prover_v2} and Kimina \citep{kimina} attempt to integrate reasoning and proving. They generate high-level sketches or subgoals before producing the final formal proof. However, we argue that these approaches suffer from a fundamental design flaw: by tightly coupling high-level planning and low-level proof generation within a single, monolithic model, they inadvertently shackle the model's full reasoning potential. Its "sketching" is constrained by what its integrated proving component can handle. This inherent limitation prevents them from fully leveraging the powerful informal reasoning capabilities demonstrated on benchmarks like PutnamBench. This, in turn, explains why even these advanced systems still fail on the most difficult problems, such as post-2000 International Mathematical Olympiad (IMO) problems.



We identify the root cause of this failure in the prevailing training paradigm: reinforcement learning with verifiable rewards (RLVR). This methodology, used to train models like DeepSeek-Prover-v2 and Kimina, rewards only the final binary success or failure of the generated Lean code. This paradigm is fundamentally misaligned with the goal of bridging the reasoning-proving gap. Instead of rewarding hard-to-define, human-like strategies (the kind that achieve >80\% informal success), the RLVR teaches a degenerated policy to maximize reward by any means necessary. It is incentivized to suppress its powerful, latent reasoning abilities in favor of heuristically decomposing goals into trivial sub-problems that can be solved by brute-forcing automatic tactics like ring, omega, or try. This over-reliance is not merely a shortcut, but a symptom of a training-induced degradation of its reasoning capabilities, effectively preventing it from generating the insightful decompositions required for truly difficult problems.

\begin{figure}
    \centering
    \includegraphics[width=1.0\linewidth]{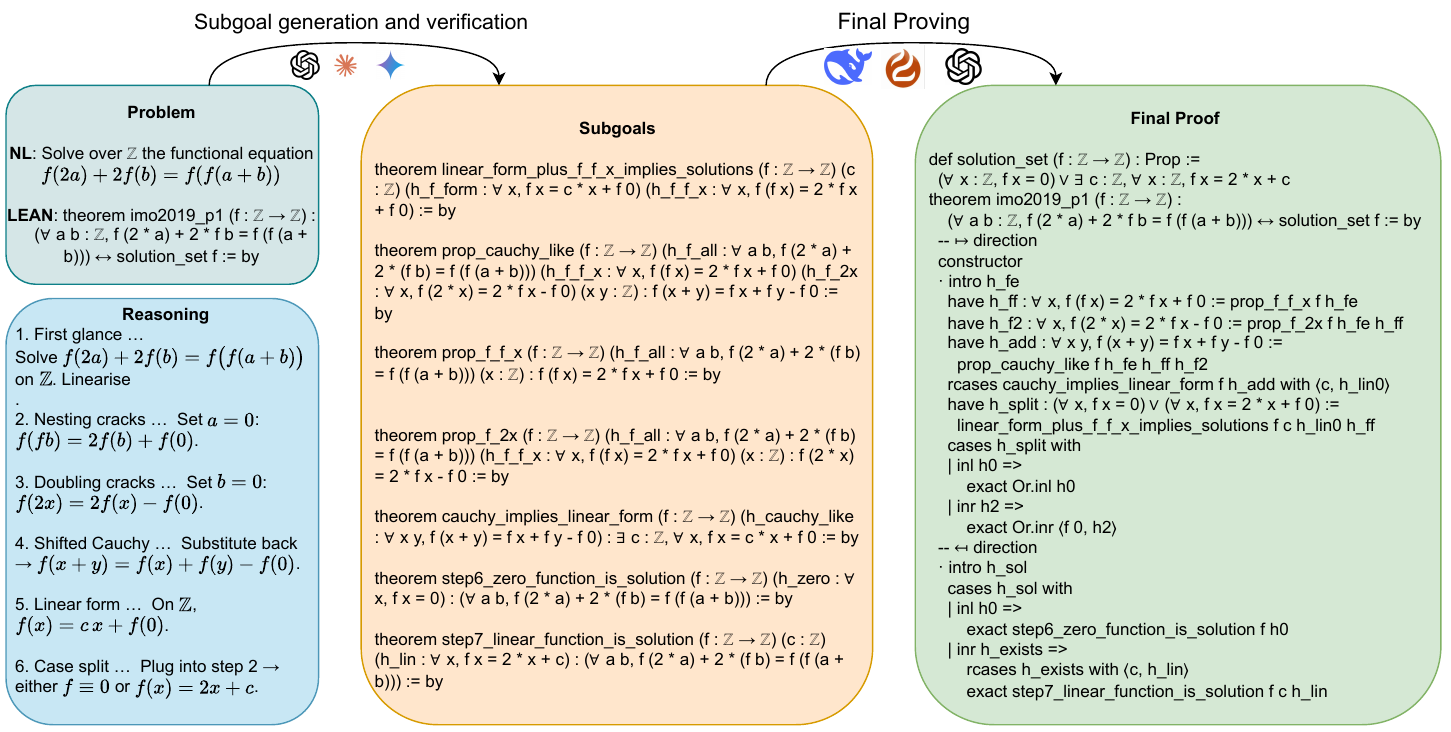}
    \caption{The overall pipeline of DRP-IMO taking the problem of IMO 2019 P1 as an example. Detailed proofs of the subgoals are omitted for brevity.}
    \label{fig:pipeline}
\end{figure}

Based on this diagnosis, we propose a new framework built on a foundational principle: the decoupling of high-level reasoning from low-level proof generation. We argue that to bridge the chasm between informal insight and formal correctness, the two processes must be handled by distinct, specialized models scheduled with greater flexibility. Our approach leverages a powerful, general-purpose LLM as a dedicated \emph{Reasoner} and a separate, efficient ATP model as the \emph{Prover}. This architectural separation liberates the \emph{Reasoner} to focus exclusively on its strength—generating high-level mathematical insights and strategic decompositions—while the Prover ensures that the resulting strategy can be grounded in formal, verifiable logic. It is this principled, decoupled bridge that we believe can finally harness the reasoning power of LLMs for formal theorem proving.

Our simple yet effective pipeline is illustrated in Figure~\ref{fig:pipeline}. Given a theorem, the Reasoner is first invoked to propose potentially useful lemmas (subgoals), expressed only as formal statements, which act as a bridge between high-level strategy and formal proof. A Prover module then attempts to verify these proposed lemmas, filtering out any that are unprovable. Finally, the Prover tackles the main theorem, now armed with a set of verified lemmas that guide the proof search and significantly reduce its complexity. This contrasts with existing approaches that follow a rigid, one-shot "reasoning-then-proving" trajectory within a single model.

We evaluate our approach on a challenging set of non-geometry IMO problems from 2000 to 2024. To the best of our knowledge, no existing open-source automated theorem prover has reported success on any problem from this set. In the experiemnts, our method successfully solves 5 of these problems: IMO 2000 Problem 2, IMO 2005 Problem 3, IMO 2011 Problem 3, IMO 2019 Problem 1, and IMO 2020 Problem 2.

\subsection*{Data Release}

To foster further research and collaboration within both the mathematics and ATP communities, we are releasing a comprehensive dataset and a project website. While our framework successfully solved 5 IMO problems, our efforts in subgoal generation and verification have yielded a much larger collection of high-quality, formally verified lemmas for a broad range of post-2000 IMO problems. We believe this resource serves a dual purpose:

\begin{itemize}[leftmargin=2em]
    \item For mathematicians and IMO researchers, this collection of machine-generated lemmas may offer novel perspectives or reveal non-obvious decompositions, potentially inspiring new human-led proof strategies.
    \item For the ATP community, our dataset acts as a new, challenging benchmark. By providing verified intermediate steps, it allows researchers to focus on solving the remaining difficult lemmas or on the final, complex proof-synthesis stage for problems currently beyond reach.
\end{itemize}

The dataset is publicly available on HuggingFace, and we are committed to its active maintenance and expansion. We welcome community contributions, such as new proofs for existing lemmas or alternative strategic decompositions. The project website, which provides access to the data repository, tracks our ongoing progress, and presents detailed case studies, can be found at \url{https://tencent-imo.github.io/}.

\begin{figure}[t]
\centering
\begin{minted}[fontsize=\small, breaklines, linenos, bgcolor=gray!10]{lean}
import Mathlib
import Aesop
set_option maxHeartbeats 0
open BigOperators Real Nat Topology Rat

theorem amc12b_2002_p7 (a b c : ℕ) (h₀ : 0 < a ∧ 0 < b ∧ 0 < c) (h₁ : b = a + 1) (h₂ : c = b + 1) (h₃ : a * b * c = 8 * (a + b + c)) : a ^ 2 + (b ^ 2 + c ^ 2) = 77 := by
    have hb : b = a + 1 := h₁
    have hc : c = a + 2 := by omega
    rw [hb, hc] at h₃
    have h4 : a = 4 := by
        have h_pos : 0 < a := h₀.left
        have : a ≤ 6 := by
            nlinarith [h₃, mul_pos h_pos (show 0 < a + 1 by omega), mul_pos h_pos (show 0 < a + 2 by omega),
                show 0 < a + 1 by omega, show 0 < a + 2 by omega]
        interval_cases a <;> omega
    have ha : a = 4 := h4
    have hb' : b = 5 := by omega
    have hc' : c = 6 := by omega
    rw [ha, hb', hc']
    norm_num
\end{minted}


\caption{Kimina solutions to amc12b\_2002\_p7. The have statements \texttt{hb} and \texttt{hc} are trivially extended from provided conditions, and statement \texttt{h4} makes a luckily guess that directly leads to the solution. Automatic tactic \texttt{omega} plays a central role throughout the proof.}
\label{fig:bad_case}
\end{figure}

\section{Related Work}

The application of Large Language Models (LLMs) to Automated Theorem Proving has evolved rapidly. Early and some recent approaches leverage the powerful sequence modeling capabilities of LLMs to generate entire formal proofs in a single, end-to-end pass. For instance, Baldur \citep{first2023baldur} generates proofs for Isabelle and incorporates a repair mechanism that learns from compiler feedback to correct flawed proofs. Other works, while still operating within a largely monolithic framework, introduce internal structure. POETRY \citep{wang2024proving} employs a recursive decomposition strategy to break down complex theorems, and LEGO-Prover \citep{wang2024legoprover} hierarchically proves and reuses lemmas to manage intermediate results within its generation process. These methods treat proof generation as a sophisticated, structured sequence generation task. In contrast, our work argues that coupling high-level reasoning and low-level proof formalization within a single model limits their potential, and we instead advocate for their explicit separation.

Recognizing the limitations of direct generation, a significant line of research has focused on integrating high-level planning or sketching, mimicking human problem-solving strategies. These methods often generate a natural language plan or a structured sketch before producing the final proof. Kimina-Prover \citep{kimina} achieves strong results by generating structured reasoning patterns prior to the formal proof. Similarly, DeepSeek-Prover-V2 \citep{ds_prover_v2}, the current state-of-the-art, integrates Chain-of-Thought (CoT) style reasoning to guide its recursive subgoal decomposition pipeline. While these methods represent a conceptual step towards our approach by acknowledging the importance of planning, they still tightly couple the planning and proving phases within a single model and a fixed workflow. Our method fundamentally differs by decoupling these two stages into distinct, specialized models, allowing for more flexible and powerful interaction, such as iterative refinement of lemmas before the final proof attempt. 

Our work builds upon the high-level philosophy of hierarchical proof generation, sharing conceptual similarities with prior efforts like Draft, Sketch, Prove \citep{jiang2023draft}, LEGO-Prover \citep{wang2024legoprover}, POETRY \citep{wang2024proving}, and Subgoal-XL \citep{zhao2024subgoalxl}. The most closely related is Draft, Sketch, Prove \citep{jiang2023draft}, which also employs a multi-stage pipeline: an LLM first drafts an informal proof, an autoformalizer then translates this draft into a formal sketch, and finally, an external prover completes the proof.

Despite this architectural resemblance, our approach makes a critical design choice that diverges significantly. Instead of attempting to autoformalize an entire unstructured natural language proof—a process that is itself a major research challenge and prone to semantic errors—we task our specialized Reasoner model with a more constrained and impactful objective: generating a diverse set of formal subgoal statements (lemmas). This design offers two key advantages. First, by focusing on generating strategic lemmas rather than full proof steps, we directly leverage the abstract reasoning strength of powerful LLMs to perform creative and non-trivial problem decomposition, which is essential for solving complex problems like those in the IMO competition. Second, by generating formal statements directly and leaving the proof generation to a dedicated Prover, we entirely bypass the fragile and error-prone autoformalization step. This ensures that the bridge between high-level reasoning and formal proving is both robust and precise.

\section{A Framework for Decoupled Reasoning and Proving}
\label{sec:method}

Our methodology is founded on the principle of decoupling high-level strategic reasoning from low-level formal proof generation. This separation allows us to use the best tool for each task: a powerful, general-purpose reasoning model for strategic decomposition, and a specialized, efficient theorem prover for formal verification. The overall workflow, illustrated in Figure~\ref{fig:pipeline}, consists of three main stages: subgoal generation, subgoal filtering, and final proof construction.

\subsection{Stage 1: Strategic Subgoal Generation with a Reasoner}
\label{sec:subgoal_gen}

The first stage aims to replicate the most crucial aspect of human mathematical problem-solving: identifying strategic intermediate steps or lemmas. For IMO-level problems, which can require hundreds of proof steps in a formal language like Lean, a brute-force search is intractable. A well-chosen set of lemmas can dramatically prune this search space by decomposing the primary goal into more manageable components.

Our central insight is to leverage a powerful, general-purpose Large Language Model as a dedicated \emph{Reasoner}, whose sole responsibility is to generate these strategic decompositions. Unlike specialized ATP models trained primarily on code, our \emph{Reasoners} (e.g., OpenAI-o3, Gemini 2.5, Claude 4 Opus) excel at high-level, semantic understanding and creative problem-solving. We task the Reasoner with generating only the *formal statements of potential lemmas, without their proofs. This deliberate constraint focuses the model on its core strength—strategic thinking—while avoiding the complexities and potential errors of full proof generation.

After evaluating several state-of-the-art models, including OpenAI-o3, Claude 4 Opus and Gemini 2.5 Pro, we selected Gemini 2.5 Pro for its superior ability to generate diverse and logically sound subgoals for complex mathematical problems. We use the following prompt, which is designed to encourage deep reasoning before outputting structured Lean statements:

\begin{tcolorbox}[
breakable,title=Prompt for Subgoal Generation, boxrule=0.5pt, colframe=black
]
You are given a very challenging theorem written in Lean 4. This theorem is too difficult to prove directly. Your task is to think step-by-step to devise a feasible and complete proof strategy for the theorem, and then decompose the original theorem into a sequence of smaller, logically coherent sub-theorems, each of which can be proved more easily.\\

Important instructions:\\

First, reason through and construct a valid and complete proof strategy for the original theorem.\\

After the solution path is clear, divide it into intermediate proof steps. Each step should be expressed as a separate sub-theorem in Lean 4, following the same syntactic and semantic format as the original theorem.\\

The decomposition should reflect deep understanding of the overall proof structure. Avoid trivial splits such as case analysis or mechanical “divide into two cases” tactics unless they are genuinely part of the reasoning process.\\

Each sub-theorem must represent a meaningful proof milestone — essentially a condensed logical step from the overall proof strategy.\\

The sub-theorems should be self-contained and provable, and collectively they should imply the original theorem.\\

Output format:\\

A brief explanation of your proof strategy (in natural language or Lean comments).\\

A list of Lean 4 theorem declarations, each representing a sub-theorem, all starting with 'theorem XXX' and ending with ':= by sorry'.\\

Ensure all sub-theorems are expressed using the same formal syntax and conventions as the input theorem.\\

Input Theorem:
\end{tcolorbox}

To reliably extract the generated lemma statements from the model's free-form text output, we apply a simple yet robust regular expression. This ensures a clean handoff to the next stage of the pipeline.

\begin{tcolorbox}[
title=Regular Expression for Lemma Extraction, boxrule=0.5pt, colframe=black
]
theorem (.*?):= by sorry
\end{tcolorbox}

\subsection{Stage 2: Subgoal Verification and Filtering}
\label{sec:subgoal_prove}

The second stage acts as a critical filter to ensure that only logically sound and verifiable lemmas are passed to the final stage. This verification step is essential for the overall soundness of our framework. For this task, we employ a dedicated \emph{Prover} model. The key requirement for this \emph{Prover} is not high-level reasoning, but rather strong tactical execution and efficiency in proving well-defined, modular goals. We selected DeepSeek-Prover-v2 (7B, CoT version) for this role, as it offers a state-of-the-art balance of proof success rate and computational efficiency. Each candidate lemma generated in Stage 1 is treated as an independent theorem. The  \emph{Prover} attempts to find a proof for it, generating up to $k$ candidate proofs. A lemma is considered "verified" and retained if at least one attempt succeeds.

This filtering process is not merely a correctness check; it is a core component of our decoupled strategy. It allows the \emph{Reasoner} in Stage 1 to be "creative" and even "speculative," proposing diverse ideas without the immediate burden of provability. The  \emph{Prover} then grounds this creativity in formal rigor, effectively selecting the most promising pathways. We set $k$ (the number of proof candidates) to 128, a value chosen empirically to balance the exploration breadth against computational cost.

\subsection{Stage 3: Final Proof Construction}
\label{sec:final_prove}

In the final stage, the prover attempts to solve the main theorem by leveraging the set of verified lemmas obtained in Stage 2. These lemmas are prepended to the context of the main problem statement, effectively enriching the problem statement with a suite of pre-proven, reusable components. This process transforms the original, monolithic proof task into a more tractable task of assembling a final proof using these intermediate results as foundational building blocks, as illustrated in Figure~\ref{fig:pipeline}.

In the final stage, the prover tries to tackle the main theorem with the aid of the set of verified lemmas from Stage 2. These lemmas are prepended to the context of the main problem statement, effectively enriching the problem with powerful, pre-proven tools. This transforms the original, monolithic challenge into a simpler task of assembling a final proof using these new building blocks, as illustrated in Figure \ref{fig:pipeline}.

A crucial challenge we identified at this stage was domain shift. We initially hypothesized that the same  \emph{Prover} from Stage 2 (DeepSeek-Prover-v2) would be optimal. However, we observed that this model, when presented with auxiliary lemmas, often struggled to effectively utilize them, likely because its training data did not emphasize this specific pattern of formal proving. It tended to ignore the provided lemmas and attempt to prove the theorem from scratch, defeating the purpose of our pipeline.

This led to a key insight: the ability to leverage existing lemmas is a distinct skill that not all provers possess equally. After further experimentation, we found that two powerful reasoner models, OpenAI-o3 and Gemini 2.5 Pro, are significantly more adept at integrating and applying the provided lemmas to construct the final proof. This highlights the importance of selecting the right tool for each sub-task in a decoupled system. While we use those reasoners for our final results, a promising direction for future work is to fine-tune specialized provers like DeepSeek-Prover-v2 on a curriculum of problems that explicitly require the use of given lemmas.

\section{Experiment}
\label{sec:exp}

We evaluate our approach on challenging problems from the International Mathematical Olympiad (IMO), focusing on non-geometry problems from the years 2000 to 2024.
Each annual IMO consists of six problems, typically including one geometry problem.
We focus on the non-geometry ones, and our method successfully solves 5 of these problems: IMO 2000 Problem 2, IMO 2005 Problem 3, IMO 2011 Problem 3, IMO 2019 Problem 1, and IMO 2020 Problem 2. We list detailed proofs in Appendix \ref{sec:solved} and show current progress on IMO 2024 problems in Appendix \ref{sec:imo_2024}.


\subsection{Comparing Reasoning Quality in IMO 2019 Problem 1}

To evaluate the qualitative difference between reasoning strategies, we analyze how our framework's \textbf{Reasoner} compares against existing prover-driven approaches when applied to a challenging benchmark: IMO 2019 Problem 1. This problem asks to find all functions $f: \mathbb{Z} \to \mathbb{Z}$ satisfying $f(2a) + 2f(b) = f(f(a+b))$ for all integers $a, b$.

Our goal is to demonstrate that the reasoning path generated by our decoupled Reasoner-Prover framework leads to a principled, structured solution strategy, in stark contrast to prover-only models, which often exhibit brittle or degenerate behavior.

\subsubsection{Our Reasoner's Strategic Decomposition}

In our framework, the Reasoner is responsible for identifying high-level mathematical structure and generating a roadmap of lemmas. On this problem, the Reasoner produces the following structured decomposition:

\begin{enumerate}
\item \textbf{Identify fundamental properties}: By strategic instantiation of the functional equation, the Reasoner isolates key identities:
\begin{itemize}
\item \texttt{prop\_f\_f\_x}: $f(f(x)) = 2f(x) + f(0)$ for all $x$
\item \texttt{prop\_f\_2x}: $f(2x) = 2f(x) - f(0)$ for all $x$
\end{itemize}

\item \textbf{Uncover additive structure}: Combining the above, the Reasoner deduces:
\begin{itemize}
\item \texttt{prop\_cauchy\_like}: $f(x+y) = f(x) + f(y) - f(0)$
\end{itemize}

\item \textbf{Characterize the function form}: Using the Cauchy-like identity, the Reasoner infers:
\begin{itemize}
\item \texttt{cauchy\_implies\_linear\_form}: There exists $c \in \mathbb{Z}$ such that $f(x) = cx + f(0)$
\end{itemize}

\item \textbf{Constrain the parameters}: Plugging this linear form into \texttt{prop\_f\_f\_x}, the Reasoner derives:
\begin{itemize}
\item \texttt{linear\_form\_plus\_f\_f\_x\_implies\_solutions}: $c$ must be either $0$ or $2$
\end{itemize}

\item \textbf{Verify candidate solutions}: Both resulting forms, $f(x) = 0$ and $f(x) = 2x + c$, are verified to satisfy the original equation.
\end{enumerate}

This decomposition exhibits genuine mathematical insight: it identifies the functional equation’s additive structure, abstracts useful intermediate results, and uses them to constrain the solution space efficiently and interpretably.

We contrast this with the behavior of current state-of-the-art prover models. Specifically, we sampled three solution attempts from the strongest publicly available model, DeepSeek Prover v2 671B \citep{ds_prover_v2}. These are representative of the general behavior we observed. For brevity, we include only partial code excerpts.

The first attempt relies on a brute-force enumeration of equations. The model instantiates the functional equation on dozens of inputs, creating a large flat pool of algebraic identities, and then invokes tactics such as \texttt{ring\_nf} and \texttt{linarith} in hopes of simplification. There is no effort to identify structure or extract reusable intermediate results. The tactic application is purely local and mechanical:

\begin{minted}[fontsize=\small, breaklines, bgcolor=gray!10]{lean}
have h₂ := hf 0 0
have h₃ := hf 0 x
have h₄ := hf x 0
have h₅ := hf x (-x)
...
have h₂₆ := hf (x + x) (-x)
ring_nf at h₂ h₃ h₄ ... h₂₆ ⊢
\end{minted}

In the second attempt, the prover tries to assert the final form of the solution—namely $f(x) = 2x + f(0)$—without having established why $f$ must be linear or what motivates such a guess. It implicitly assumes the desired conclusion and attempts to work backward through aggressive simplification. This reveals a logical gap: the model never proves the Cauchy-like identity nor justifies why a linear form should even be expected.

\begin{minted}[fontsize=\small, breaklines, bgcolor=gray!10]{lean}
have h₂₉ : f x = 2 * x + f 0 := by
  have h₂₉₁ := hf x 0
  ...
  ring_nf at h₂₉₁ h₂₉₂ ... ⊢
  <;> linarith
\end{minted}

The third attempt generates an even larger collection of equation instances, trying all possible combinations of inputs into the original functional equation, and then offloads the burden of reasoning onto a generic decision procedure like \texttt{omega}. Again, no insight is gained; the solution depends entirely on the capacity of low-level tactics to blindly traverse the search space.

\begin{minted}[fontsize=\small, breaklines, bgcolor=gray!10]{lean}
have h₃₁ : f x = 2 * x + (f 0 - 2 * 0) := by
  have h₃₂ := hf 0 0
  ...
  have h₄₂ := hf 1 (x - 1)
  ring_nf at h₃₂ h₃₃ ... h₄₂ ⊢
  omega
\end{minted}
These degenerate strategies are a direct consequence of the reward signals guiding the training of prover models. When models are rewarded solely for producing verifiable proofs, they learn to exploit patterns that maximize verification success, not reasoning quality. Brute-force instantiation followed by tactic chains often suffices on simple benchmarks, so models internalize that behavior—even when such strategies fail to scale to Olympiad-level problems. In these more complex settings, the search space is too vast, and the necessary structural insights (such as recognizing the shifted Cauchy identity) cannot be discovered by purely local manipulations.

In contrast, our framework deliberately separates the high-level reasoning process from low-level proof verification. The Reasoner is not constrained by the demands of tactic execution or code generation; it operates at the level of abstraction and mathematical insight. By generating a chain of semantically meaningful lemmas, it defines a proof skeleton that guides the Prover and drastically reduces the search space. This separation enables the kind of reasoning that mirrors how human mathematicians approach challenging problems: by detecting invariants, proposing transformations, and narrowing the solution space through conceptual understanding. As this case study illustrates, this leads to reasoning that is not only verifiable, but also interpretable, reusable, and robust.


\subsection{Degradation of Mathematical Reasoning in Specialised Provers}
\label{sec:reasoning_degradation}

\textbf{Motivation.} A central hypothesis of our work is that the prevailing reinforcement learning with verifiable rewards (RLVR) paradigm, while effective for optimizing success rates on specific ATP benchmarks, may inadvertently cause a degradation in the model's intrinsic mathematical reasoning capabilities. The reward signal, being solely dependent on the formal proof's success, does not explicitly value the quality or correctness of the natural language reasoning that may precede it. To test this hypothesis, we designed an experiment to isolate and measure this potential degradation.

\textbf{Experimental Setup.} We compare the performance of a specialized prover model with its general-purpose base model on standard mathematical reasoning benchmarks that do not involve formal proof generation. Specifically, we selected:
\begin{itemize}
    \item \textbf{Base Model:} \texttt{Qwen2.5-Math-7B-Instruct}, the foundational model upon which the Kimina-Prover is built, which is highly capable in general mathematical problem-solving.
    \item \textbf{Prover:} \texttt{Kimina-Prover-Preview-Distill-7B}, a state-of-the-art prover initialized from \texttt{Qwen2.5-Math-7B-Instruct} and fine-tuned for Lean-based theorem proving.
\end{itemize}
We evaluated both models on the \textbf{MATH} and \textbf{AIME} benchmarks. We found that with appropriate prompting, \texttt{Kimina-Prover} can still generate high-quality solutions to math problems, rather than generating Lean code. This allowed for a direct comparison of their problem-solving accuracy. We report the pass@k accuracy for both models on AIME24.

\begin{table}[h!]
\centering
\caption{Performance comparison on general mathematical reasoning benchmarks.}
\label{tab:reasoning_degradation}
\resizebox{\linewidth}{!}{%
\begin{tabular}{@{}lccccc@{}}
\toprule
 & \textbf{MATH} & \multicolumn{4}{c}{\textbf{AIME24}} \\
\cmidrule(lr){2-2} \cmidrule(lr){3-6}
\textbf{Model} & \textbf{pass@1} & \textbf{pass@1} & \textbf{pass@4} & \textbf{pass@8} & \textbf{pass@16} \\ \midrule
\texttt{Qwen2.5-Math-7B-Instruct} (base model) & \textbf{83.6\%} & \textbf{16.7\%} & \textbf{33.3\%} & \textbf{43.3\%} & \textbf{46.7\%} \\
\texttt{Kimina-Prover-Preview-Distill-7B} (prover) & 78.7\% & 11.0\% & 24.1\% & 32.0\% & 40.9\% \\ \midrule
\textbf{Performance Drop (pts)} & \textbf{-4.9} & \textbf{-5.7} & \textbf{-9.2} & \textbf{-11.3} & \textbf{-5.8} \\ \bottomrule
\end{tabular}%
}
\end{table}

The results, presented in Table~\ref{tab:reasoning_degradation}, provide clear evidence supporting our hypothesis. On both benchmarks, the \texttt{Kimina-Prover} exhibits a marked decline in performance compared to its base model. The single-attempt accuracy (pass@1) drops by 4.9 percentage points on MATH and 5.7 points on AIME. Crucially, this performance gap is not an isolated phenomenon but persists robustly across multi-sample evaluations on the challenging AIME dataset. The performance delta remains substantial at pass@4, widens further at pass@8, and is still significant at pass@16. This confirms that the specialization process for formal theorem proving, while boosting performance on ATP tasks, comes at the cost of broader mathematical reasoning skills. This finding strongly motivates our decoupled approach: instead of attempting to force a single model to excel at both high-level reasoning and low-level formalization, we should leverage a dedicated, un-degraded reasoning model for the former, preserving its full intellectual capacity.

\subsection{Further Discussions}

\paragraph{On the Utilization of External Knowledge: Lemmas vs. \texttt{have} Statements.}

A critical challenge we encountered during the development of our pipeline was the effective integration of verified subgoals into the final proof stage. We observed a significant behavioral pattern: when verified subgoals were provided as standalone \texttt{lemma}s in the context, many state-of-the-art provers, including DeepSeek-Prover-v2, tended to ignore them. Instead of leveraging these pre-proven facts, the models often attempted to prove the main theorem from scratch, indicating a form of "contextual blindness" or a bias towards self-contained proof generation learned during their fine-tuning.

The "contextual blindness" phenomenon contrasts with using a \texttt{have} statement within a proof, which forces the model to work with a specific, locally-defined fact \citep{ds_prover_v2,cao2025reviving}. However, \texttt{have} statements require strict alignment of symbols and definitions with the current proof state, limiting their flexibility. Standalone lemmas, in theory, offer a more powerful and flexible mechanism for incorporating external knowledge, as they do not impose such rigid constraints. This flexibility is crucial for solving complex problems where reusing established results is key. Our findings suggest that a significant gap exists in the ability of current provers to effectively utilize modular, pre-proven knowledge. A crucial direction for future work is therefore to develop or fine-tune provers to specifically excel at this "proof continuation" task, enabling them to robustly accept and leverage a library of existing theorems and lemmas.

\paragraph{Limitations and Failure Analysis.}

Our analysis on unsolved statements reveals two primary bottlenecks in the current pipeline, highlighting the remaining gap between our automated system and human-level mathematical ingenuity.

First, the primary mode of failure is the Prover's inability to verify critical, complex lemmas. For many unsolved problems, our Reasoner successfully identified a plausible high-level strategy, but the constituent lemmas were simply too difficult for the Prover module to handle. To validate this, we conducted an oracle experiment where we manually proved these bottleneck lemmas (or replaced their proofs with \texttt{sorry}). In this idealized setting, our framework was able to solve a significantly larger number of IMO problems. This demonstrates that the performance of the entire pipeline is currently bounded by the raw theorem-proving power of the Prover component.
Another potential approach involves further decomposing unresolved lemmas into simpler sub-lemmas to facilitate their proof.

Second, we identified a fundamental difference in the reasoning style of our Reasoner compared to human mathematicians. Through manual inspection of our Reasoner's outputs against official IMO solutions, we observed that human proofs often hinge on a single, "magical" insight or a clever re-framing of the problem that dramatically simplifies the proof. Our Reasoner, reflecting its LLM nature, excels at systematic, step-by-step decomposition and logical deduction. It is proficient at breaking a problem down into a clear chain of sub-problems but struggles to generate the kind of non-obvious, highly creative leaps that characterize elegant human solutions. This "ingenuity gap" represents a deeper, more fundamental challenge for LLM-based reasoning and is a key bottleneck for our framework on the most difficult problems. 

\section{Conclusion}
\label{sec:conclusion}
In this work, we introduce a novel framework for automated theorem proving that addresses a core limitation in current systems: the degradation of mathematical reasoning ability caused by end-to-end RLVR training. Our central contribution is the principle of decoupling strategic reasoning from formal proof generation. We achieve this by delegating high-level strategic thinking to a dedicated Reasoner—a powerful, general-purpose LLM whose nuanced reasoning capabilities are often compromised during specialized prover fine-tuning. This Reasoner formulates its strategy as a set of formal subgoal lemmas, providing a more general and powerful mechanism for problem decomposition than restrictive in-proof statements. This modular design ensures proof search is guided by a coherent, human-like plan.

Our evaluation on a challenging set of post-2000 IMO problems, where we successfully solved 5 problems previously unsolved by any open-source prover, provides strong evidence for the efficacy of our decoupled approach. We acknowledge that our current evaluation is limited and that the multi-stage nature of our pipeline presents scalability challenges. A key priority for future work is to streamline this pipeline to enable comprehensive evaluation on large-scale benchmarks such as miniF2F and ProofNet. Ultimately, we believe that by separating the art of strategy from the science of verification, our work paves the way for more robust, scalable, and insightful automated reasoning systems capable of tackling the frontiers of mathematics.

\bibliography{colm2024_conference}
\bibliographystyle{colm2024_conference}

\appendix

\section{Case Studies on IMO 2024 Problems}
\label{sec:imo_2024}

This section provides a detailed analysis of our framework's progress on two problems from the IMO 2024. For each problem, we present the main theorem, summarize the key sub-theorems (lemmas) that our framework successfully generated and proved, and identify the critical remaining steps required to complete the full proof.

\subsection{Analysis of IMO 2024, Problem 1}

\subsubsection{Main Theorem}
The problem asks to prove the equivalence between a real number \(a\) being an even integer and a specific divisibility property holding for all positive integers \(n\).

\begin{minted}[fontsize=\small, breaklines,  bgcolor=gray!10]{lean}
theorem imo2024_p1 (a : ℝ) : 
    (∃ m : ℤ, a = 2 * m) ↔ ∀ n : ℕ, 0 < n → (n : ℤ) ∣ ∑ i in Finset.Icc 1 n, ⌊i * a⌋ := by
\end{minted}

The proof naturally splits into two directions:
\begin{itemize}
    \item \textbf{Forward Direction (1) \(\rightarrow\) (2):} If \(a = 2m\) for some integer \(m\), then the divisibility property holds.
    \item \textbf{Reverse Direction (2) \(\rightarrow\) (1):} If the divisibility property holds, then \(a\) must be of the form \(2m\).
\end{itemize}

\subsubsection{Progress Summary and Key Proven Lemmas}

Our framework has made substantial progress on this problem, most notably by completely proving the forward direction and establishing the crucial strategic lemmas for the reverse direction.

\paragraph{Forward Direction: Complete.} The framework successfully proved that if \(a\) is an even integer, the divisibility property holds. This was accomplished through several lemmas, culminating in a direct proof of the implication.

\begin{minted}[fontsize=\small, breaklines,  bgcolor=gray!10]{lean}
-- Proved: The forward implication of the main theorem.
theorem imo2024_p1_forward_implication (a : ℝ) :
  (∃ m : ℤ, a = 2 * m) → (∀ n : ℕ, 0 < n → (n : ℤ) ∣ ∑ i in Finset.Icc 1 n, ⌊i * a⌋) := by
\end{minted}

\paragraph{Reverse Direction: Key Strategic Lemmas Proven.} For the more challenging reverse direction, our system proved two cornerstone lemmas that are essential to the standard human solution strategy.

\begin{enumerate}
    \item \textbf{Periodicity of the Condition:} The framework proved that the divisibility property is periodic with a period of any even integer. This is a powerful strategic result, as it allows reducing the problem for any real number \(a\) to an equivalent problem for a number in a bounded interval (e.g., \([0, 2]\)).
    
    \begin{minted}[fontsize=\small, breaklines,  bgcolor=gray!10]{lean}
    -- Proved: The divisibility property is periodic by any even integer.
    theorem divisibility_is_periodic_by_even_integers (a : ℝ) (m : ℤ) :
        (∀ n : ℕ, 0 < n → (n : ℤ) ∣ ∑ i in Finset.Icc 1 n, ⌊i * a⌋) ↔
        (∀ n : ℕ, 0 < n → (n : ℤ) ∣ ∑ i in Finset.Icc 1 n, ⌊i * (a - 2 * m)⌋) := by
    \end{minted}
    
    \item \textbf{Special Case for Integers:} The system proved that if \(a\) is an integer satisfying the divisibility property, it must be an even integer. This fully resolves the reverse direction for the specific case where \(a \in \mathbb{Z}\).
    
    \begin{minted}[fontsize=\small, breaklines,  bgcolor=gray!10]{lean}
    -- Proved: An integer satisfying the property must be even.
    theorem integer_must_be_even (a : ℤ)
        (h_div_int : ∀ n : ℕ, 0 < n → (n : ℤ) ∣ ∑ i in Finset.Icc 1 n, ⌊(i : ℝ) * (a : ℝ)⌋) :
        Even a := by
    \end{minted}
\end{enumerate}

\subsubsection{Analysis of the Remaining Proof Goal}

With the forward direction complete and the key periodicity lemma established, the entire proof now hinges on a single, final sub-problem.

\paragraph{The Final Step: Proving the Base Case for the Periodicity.}
The established lemmas allow us to reason as follows: Assume a real number \(a\) satisfies the divisibility property. We can find an integer \(m\) such that \(a' = a - 2m\) lies in the interval \([-1, 1]\). Due to the proven periodicity, \(a'\) must also satisfy the divisibility property. If we can prove that the only number in \([-1, 1]\) satisfying the property is \(0\), it would imply \(a' = 0\), which means \(a = 2m\), completing the proof.

Therefore, the critical missing lemma is to show that for any number \(a \in [-1, 1]\) (or a similar interval like \((-1, 1]\)), if it satisfies the property, it must be zero.

\begin{tcolorbox}[colback=red!5!white,colframe=red!75!black,title=Critical Missing Lemma]
\textbf{Goal:} To prove that if a real number \(a\) in the interval \([-1, 1]\) satisfies the universal divisibility condition, then \(a\) must be \(0\).
\begin{minted}[fontsize=\small, breaklines, bgcolor=gray!10]{lean}
theorem univ_divisibility_in_interval_implies_zero (a : ℝ) (ha_bound : a ∈ Set.Icc (-1) 1)
    (h_prop : ∀ n : ℕ, 0 < n → (n : ℤ) ∣ ∑ i in Finset.Icc 1 n, ⌊i * a⌋) :
    a = 0 := by
\end{minted}
\end{tcolorbox}

Successfully proving this final lemma would allow us to connect all the previously established results and formally complete the entire proof for IMO 2024, Problem 1. Our framework has successfully navigated the problem to its final, decisive step.

\subsection{Analysis of IMO 2024, Problem 2}

\subsubsection{Main Theorem}
This problem concerns a property of the greatest common divisor (GCD) of two exponential sequences. It asks to prove that the GCD becoming constant for all sufficiently large \(n\) is equivalent to \(a\) and \(b\) both being 1.

\begin{minted}[fontsize=\small, breaklines,  bgcolor=gray!10]{lean}
theorem imo2024_p2 (a b : ℕ+) : 
    (a, b) = (1, 1) ↔ ∃ g N : ℕ+, ∀ n : ℕ, N ≤ n → Nat.gcd (a^n + b) (b^n + a) = g := by
\end{minted}

The proof structure involves a straightforward forward direction and a more complex reverse direction, which is typically solved by considering cases.

\subsubsection{Progress Summary and Key Proven Lemmas}

Our framework successfully proved the simple forward direction and made significant headway on the reverse direction by proving the special case where \(a=b\).

\paragraph{Forward Direction: Complete.} The framework easily proved that if \(a=1\) and \(b=1\), the GCD sequence is constant. In this case, \(\gcd(1^n+1, 1^n+1) = \gcd(2,2) = 2\) for all \(n\), so one can choose \(g=2\) and \(N=1\).

\begin{minted}[fontsize=\small, breaklines,  bgcolor=gray!10]{lean}
-- Proved: The forward implication of the main theorem.
theorem imo2024_p2_forward_implication (a b : ℕ+) :
  (a, b) = (1, 1) → ∃ g N : ℕ+, ∀ n : ℕ, N ≤ n → Nat.gcd (a^n + b) (b^n + a) = g := by
\end{minted}

\paragraph{Reverse Direction: Special Case \(a=b\) Proven.} For the reverse direction, the framework identified and fully proved the crucial sub-case where \(a=b\). It correctly deduced that if \(a=b\) and the GCD property holds, then \(a\) must be 1. The reasoning relies on the fact that if \(a=b\), the GCD is \(\gcd(a^n+a, a^n+a) = a^n+a\). For this sequence to be constant for \(n \geq N\), \(a\) cannot be greater than 1.

\begin{minted}[fontsize=\small, breaklines,  bgcolor=gray!10]{lean}
-- Proved: If a=b and the GCD property holds, then a=1 and b=1.
theorem imo2024_p2_bwd_a_eq_b (a b : ℕ+) (h_ab : a = b)
    (h_gcd_const : ∃ g N : ℕ+, ∀ n : ℕ, N ≤ n → Nat.gcd (a^n + b) (b^n + a) = g) :
    (a, b) = (1, 1) := by
\end{minted}

This lemma is supported by another proven sub-theorem stating that an exponential sequence like \(a^n+a\) cannot be eventually constant if \(a > 1\).

\begin{minted}[fontsize=\small, breaklines,  bgcolor=gray!10]{lean}
-- Proved: An exponential sequence is not eventually constant for a > 1.
theorem exponential_not_eventually_constant (a : ℕ+) :
  a > 1 → ¬∃ g N : ℕ+, ∀ n : ℕ, N ≤ n → a^n + a = g := by
\end{minted}

\subsubsection{Analysis of the Remaining Proof Goal}

With the forward direction and the \(a=b\) case of the reverse direction complete, the entire proof now rests on resolving the case where \(a \neq b\).

\paragraph{The Final Step: Proving the Case \(a \neq b\) Leads to a Contradiction.}
The standard human approach for the case \(a \neq b\) (without loss of generality, assume \(a > b\)) is to show that the GCD sequence, \(d_n = \gcd(a^n+b, b^n+a)\), cannot be eventually constant if \(a > 1\). A common technique involves using properties of the GCD, such as \(\gcd(X, Y) = \gcd(X, Y - kX)\). Applying this here:
\[ d_n = \gcd(a^n+b, b^n+a) = \gcd(a^n+b, b^n+a - b^{n-1}(a^n+b)) \]
which simplifies the second term. The key is to show that if \(a>b\geq 1\), this sequence cannot be constant for large \(n\).

Therefore, the critical missing lemma is to prove by contradiction that if the GCD property holds, the case \(a \neq b\) is impossible unless \(a=b=1\) (which is already covered).

\begin{tcolorbox}[colback=red!5!white,colframe=red!75!black,title=Critical Missing Lemma]
\textbf{Goal:} To prove that if \(a \neq b\), the GCD property cannot hold. A common way is to show that if \(a > b\), the GCD sequence is not constant.
\begin{minted}[fontsize=\small, breaklines, bgcolor=gray!10]{lean}
-- This lemma is stated to lead to a contradiction with the main hypothesis.
theorem gcd_is_not_eventually_constant_if_unequal (a b : ℕ+) (h_neq : a ≠ b) :
    ¬(∃ g N : ℕ+, ∀ n : ℕ, N ≤ n → Nat.gcd (a^n + b) (b^n + a) = g) := by
-- A more direct approach to prove is:
-- if a > b >= 1, then the sequence is not constant.
\end{minted}
Or, framing it to directly complete the main proof:
\begin{minted}[fontsize=\small, breaklines,  bgcolor=gray!10]{lean}
-- This lemma, combined with the a=b case, would complete the proof.
theorem p2_bwd_dir_a_neq_b (a b : ℕ+) :
  (∃ g N : ℕ+, ∀ n : ℕ, N ≤ n → Nat.gcd (a^n + b) (b^n + a) = g) → a ≠ b → False := by
\end{minted}
\end{tcolorbox}

By proving that the GCD property cannot hold for distinct positive integers \(a\) and \(b\), our framework would successfully eliminate the only remaining case, thereby completing the proof for IMO 2024, Problem 2.

\section{Solved IMO problems}
\label{sec:solved}

\subsection{IMO 2020 P2}
\begin{minted}[fontsize=\small, breaklines, linenos, bgcolor=gray!10]{lean}
-- Solution to IMO 2020 P2 by DRP-IMO

import Mathlib
import Aesop

set_option maxHeartbeats 0

open BigOperators Real Nat Topology Rat 

/--Consider four real numbers \( a, b, c, \) and \( d \) such that \( 0 < d \leq c \leq b \leq a \) and their sum is equal to 1, i.e., \( a + b + c + d = 1 \). Prove that the product of the weighted sum \( a + 2b + 3c + 4d \) and the sum of their squares \( a^2 + b^2 + c^2 + d^2 \) is less than 1.-/
theorem weighted_sum_times_sum_sq_lt_one (a b c d : ℝ) (hd_pos : 0 < d) (hdc : d ≤ c) (hcb : c ≤ b) (hba : b ≤ a) (h_sum_eq_1 : a + b + c + d = 1) :
  (a + 2*b + 3*c + 4*d) * (a^2 + b^2 + c^2 + d^2) < 1 := by
  have h_a_lt_one : a < 1 := by
    have h₁ : a + b + c + d = 1 := h_sum_eq_1
    have h₂ : 0 < b := by linarith
    have h₃ : 0 < c := by linarith
    have h₄ : 0 < d := hd_pos
    nlinarith
  
  have h_b_lt_one : b < 1 := by
    have h₁ : a ≥ b := by linarith
    have h₂ : a < 1 := h_a_lt_one
    nlinarith
  
  have h_c_lt_one : c < 1 := by
    have h₁ : b ≥ c := by linarith
    have h₂ : b < 1 := h_b_lt_one
    nlinarith
  
  have h_d_lt_one : d < 1 := by
    have h₁ : c ≥ d := by linarith
    have h₂ : c < 1 := h_c_lt_one
    nlinarith
  
  have h_a2_lt_a : a^2 < a := by
    have h₁ : a < 1 := h_a_lt_one
    have h₂ : a > 0 := by linarith
    nlinarith [sq_pos_of_pos h₂]
    <;> nlinarith
  
  have h_b2_lt_b : b^2 < b := by
    have h₁ : b < 1 := h_b_lt_one
    have h₂ : b > 0 := by linarith
    nlinarith [sq_pos_of_pos h₂]
    <;> nlinarith
  
  have h_c2_lt_c : c^2 < c := by
    have h₁ : c < 1 := h_c_lt_one
    have h₂ : c > 0 := by linarith
    nlinarith [sq_pos_of_pos h₂]
    <;> nlinarith
  
  have h_d2_lt_d : d^2 < d := by
    have h₁ : d < 1 := h_d_lt_one
    have h₂ : d > 0 := hd_pos
    nlinarith [sq_pos_of_pos h₂]
    <;> nlinarith
  
  have h_sum_sq_lt_one : a^2 + b^2 + c^2 + d^2 < 1 := by
    nlinarith [h_a2_lt_a, h_b2_lt_b, h_c2_lt_c, h_d2_lt_d]
    <;> linarith
  
  have h_main : (a + 2*b + 3*c + 4*d) * (a^2 + b^2 + c^2 + d^2) < 1 := by
    have h₁ : 0 < a + 2 * b + 3 * c + 4 * d := by
      nlinarith [hd_pos, hcb, hba, hdc, h_sum_eq_1]
    have h₂ : a ^ 2 + b ^ 2 + c ^ 2 + d ^ 2 < 1 := h_sum_sq_lt_one
    nlinarith [h₁, h₂]
    <;> nlinarith
  
  exact h_main


theorem vars_are_in_0_1 (a b c d : ℝ) (hd0 : 0 < d) (hdc : d ≤ c) (hcb : c ≤ b) (hba : b ≤ a) (h1 : a + b + c + d = 1) :
  (0 < a ∧ a < 1) ∧ (0 < b ∧ b < 1) ∧ (0 < c ∧ c < 1) ∧ (0 < d ∧ d < 1) := by
  have h_a_pos : 0 < a := by
    nlinarith [hdc, hcb, hba, hd0, h1]
    <;> nlinarith
  
  have h_a_lt_1 : a < 1 := by
    have h2 : a < 1 := by
      nlinarith [h1, h_a_pos, hba, hcb, hdc, hd0]
    exact h2
  
  have h_b_pos : 0 < b := by
    nlinarith [hdc, hcb, hba, hd0, h1]
  
  have h_b_lt_1 : b < 1 := by
    have h2 : b < 1 := by
      nlinarith [h1, h_a_pos, h_a_lt_1, hba, hcb, hdc, hd0]
    exact h2
  
  have h_c_pos : 0 < c := by
    nlinarith [hdc, hcb, hba, hd0, h1]
  
  have h_c_lt_1 : c < 1 := by
    have h2 : c < 1 := by
      nlinarith [h1, h_a_pos, h_a_lt_1, h_b_pos, h_b_lt_1, hba, hcb, hdc, hd0]
    exact h2
  
  have h_d_pos : 0 < d := by
    exact hd0
  
  have h_d_lt_1 : d < 1 := by
    have h2 : d < 1 := by
      nlinarith [h1, h_a_pos, h_a_lt_1, h_b_pos, h_b_lt_1, h_c_pos, h_c_lt_1, hdc, hcb, hba, hd0]
    exact h2
  
  refine' ⟨⟨h_a_pos, h_a_lt_1⟩, ⟨h_b_pos, h_b_lt_1⟩, ⟨h_c_pos, h_c_lt_1⟩, ⟨h_d_pos, h_d_lt_1⟩⟩


theorem imo2020_q2 (a b c d : ℝ) (hd0 : 0 < d) (hdc : d ≤ c) (hcb : c ≤ b) (hba : b ≤ a) (h1 : a + b + c + d = 1) :
    (a + 2 * b + 3 * c + 4 * d) * (a ^ a * b ^ b * c ^ c * d ^ d) < 1 := by
  -- strategy:
  -- 1. apply weighted AM-GM inequality to prove a^a * b^b * c^c * d^d ≤ a^2 + b^2 + c^2 + d^2
  -- 2. ues subgoal 'weighted_sum_times_sum_sq_lt_one' to get (a + 2*b + ...) * (a^2 + b^2 + ...) < 1
  -- 3. combine both results to reach final conclusion

  -- define S
  let S := a^2 + b^2 + c^2 + d^2

  -- step 1: apply weighted AM-GM inequality
  -- we need to prove a^a * b^b * c^c * d^d ≤ S
  have h_geom_mean_le_sum_sq : a ^ a * b ^ b * c ^ c * d ^ d ≤ S := by
    -- in order to use the subgoal 'geom_mean_le_arith_mean_weighted', we use Fin 4 as an index type
    let w : Fin 4 → ℝ := ![a, b, c, d]
    let z : Fin 4 → ℝ := ![a, b, c, d]

    -- check AM-GM prerequisite
    have h_pos_conds : (0 < a) ∧ (0 < b) ∧ (0 < c) ∧ (0 < d) := by
      have h_all := vars_are_in_0_1 a b c d hd0 hdc hcb hba h1
      exact ⟨h_all.1.1, h_all.2.1.1, h_all.2.2.1.1, h_all.2.2.2.1⟩

    -- 1. non-negative weights
    have h_weights_nonneg : ∀ i, 0 ≤ w i := by
      intro i; fin_cases i <;> simp [w] <;> linarith [h_pos_conds.1, h_pos_conds.2.1, h_pos_conds.2.2.1, h_pos_conds.2.2.2]

    -- 2. weights sum-up to 1
    have h_weights_sum_1 : ∑ i, w i = 1 := by
      simp [w, Fin.sum_univ_four, h1]

    -- 3. non-negative values
    have h_values_nonneg : ∀ i, 0 ≤ z i := by
      intro i; fin_cases i <;> simp [z] <;> linarith [h_pos_conds.1, h_pos_conds.2.1, h_pos_conds.2.2.1, h_pos_conds.2.2.2]

    -- use the subgoal based on AM-GM
    have h_am_gm := geom_mean_le_arith_mean_weighted (Finset.univ) w z (fun i _ ↦ h_weights_nonneg i) h_weights_sum_1 (fun i _ ↦ h_values_nonneg i)

    -- transform AM-GM results to the form we want
    -- `simp` will handle a*a -> a^2
    simp only [Fin.prod_univ_four, Fin.sum_univ_four, w, z, ← pow_two] at h_am_gm

    -- it will replace 'S' to 'a^2 + b^2 + c^2 + d^2'
    unfold S
    -- now the target fully matchs 'h_am_gm'
    exact h_am_gm

  -- step 2: get results from key lemmas
  have h_main_ineq : (a + 2 * b + 3 * c + 4 * d) * S < 1 := by
    exact weighted_sum_times_sum_sq_lt_one a b c d hd0 hdc hcb hba h1

  -- step 3 & 4 & 5: assumble final proof
  calc
    (a + 2*b + 3*c + 4*d) * (a^a * b^b * c^c * d^d)
    -- first, use the results from step 1, we need to prove (a + 2*b + ...) is positive
    -- lemma 'vars_are_in_0_1' guarantees a,b,c,d > 0, thus their weighted sum also > 0
    _ ≤ (a + 2*b + 3*c + 4*d) * S := by
        apply mul_le_mul_of_nonneg_left h_geom_mean_le_sum_sq
        have h_pos_conds := vars_are_in_0_1 a b c d hd0 hdc hcb hba h1
        linarith [h_pos_conds.1.1, h_pos_conds.2.1.1, h_pos_conds.2.2.1.1, h_pos_conds.2.2.2.1]
    -- then, use the results from step 2 to finish proving
    _ < 1 := h_main_ineq
\end{minted}

\subsection{IMO 2019 P1}
\begin{minted}[fontsize=\small, breaklines, linenos, bgcolor=gray!10]{lean}
-- Solution to IMO 2019 P1 by DRP-IMO

import Mathlib
import Aesop

set_option maxHeartbeats 0

open BigOperators Real Nat Topology Rat 

def solution_set (f : ℤ → ℤ) : Prop :=
  (∀ x : ℤ, f x = 0) ∨ ∃ c : ℤ, ∀ x : ℤ, f x = 2 * x + c

theorem linear_form_plus_f_f_x_implies_solutions (f : ℤ → ℤ) (c : ℤ)
    (h_f_form : ∀ x, f x = c * x + f 0) (h_f_f_x : ∀ x, f (f x) = 2 * f x + f 0) :
  (∀ x, f x = 0) ∨ (∀ x, f x = 2 * x + f 0) := by
  have h_c_squared : c^2 = 2 * c := by
    have h1 := h_f_f_x 1
    have h2 := h_f_f_x 0
    have h3 := h_f_f_x (-1)
    have h4 := h_f_form 1
    have h5 := h_f_form 0
    have h6 := h_f_form (-1)
    have h7 := h_f_form (f 1)
    have h8 := h_f_form (f 0)
    have h9 := h_f_form (f (-1))
    have h10 := h_f_f_x (f 1)
    have h11 := h_f_f_x (f 0)
    have h12 := h_f_f_x (f (-1))
    have h13 := h_f_form (c * 1 + f 0)
    have h14 := h_f_form (c * 0 + f 0)
    have h15 := h_f_form (c * (-1) + f 0)
    have h16 := h_f_form (c * (f 1) + f 0)
    have h17 := h_f_form (c * (f 0) + f 0)
    have h18 := h_f_form (c * (f (-1)) + f 0)
    ring_nf at h1 h2 h3 h4 h5 h6 h7 h8 h9 h10 h11 h12 h13 h14 h15 h16 h17 h18 ⊢
    nlinarith [sq_nonneg (c - 2), sq_nonneg (c + 2), sq_nonneg (c - 1), sq_nonneg (c + 1)]
  
  have h_c_cases : c = 0 ∨ c = 2 := by
    have h₁ : c ^ 2 = 2 * c := h_c_squared
    have h₂ : c = 0 ∨ c = 2 := by
      have h₃ : c * (c - 2) = 0 := by
        linarith
      have h₄ : c = 0 ∨ c - 2 = 0 := by
        apply eq_zero_or_eq_zero_of_mul_eq_zero h₃
      cases h₄ with
      | inl h₄ =>
        exact Or.inl h₄
      | inr h₄ =>
        have h₅ : c = 2 := by
          omega
        exact Or.inr h₅
    exact h₂
  
  have h_main : (∀ x, f x = 0) ∨ (∀ x, f x = 2 * x + f 0) := by
    cases h_c_cases with
    | inl h_c_zero =>
      -- Case c = 0
      have h_f_zero : ∀ x, f x = f 0 := by
        intro x
        have h₁ := h_f_form x
        simp [h_c_zero] at h₁ ⊢
        <;> linarith
      have h_f_zero_zero : f 0 = 0 := by
        have h₁ := h_f_f_x 0
        have h₂ := h_f_form 0
        have h₃ := h_f_form (f 0)
        have h₄ := h_f_f_x (f 0)
        simp [h_f_zero] at h₁ h₂ h₃ h₄ ⊢
        <;>
        (try omega) <;>
        (try
          {
            nlinarith [h_f_form 0, h_f_form 1, h_f_form (-1), h_f_form (f 0)]
          }) <;>
        (try
          {
            cases' h_c_cases with h_c_zero h_c_two <;> simp_all [h_c_zero, h_c_two] <;>
            (try omega) <;>
            (try nlinarith) <;>
            (try linarith)
          }) <;>
        (try
          {
            aesop
          })
      have h_f_zero_all : ∀ x, f x = 0 := by
        intro x
        have h₁ := h_f_zero x
        have h₂ := h_f_zero 0
        have h₃ := h_f_zero (-1)
        have h₄ := h_f_zero 1
        simp [h_f_zero_zero] at h₁ h₂ h₃ h₄ ⊢
        <;>
        (try omega) <;>
        (try nlinarith) <;>
        (try aesop)
        <;>
        (try
          {
            simp_all [h_f_form, h_c_zero]
            <;>
            (try omega) <;>
            (try nlinarith) <;>
            (try aesop)
          })
      exact Or.inl h_f_zero_all
    | inr h_c_two =>
      -- Case c = 2
      have h_f_form_two : ∀ x, f x = 2 * x + f 0 := by
        intro x
        have h₁ := h_f_form x
        simp [h_c_two] at h₁ ⊢
        <;> linarith
      exact Or.inr h_f_form_two
  
  exact h_main

theorem prop_cauchy_like (f : ℤ → ℤ) (h_f_all : ∀ a b, f (2 * a) + 2 * (f b) = f (f (a + b)))
    (h_f_f_x : ∀ x, f (f x) = 2 * f x + f 0) (h_f_2x : ∀ x, f (2 * x) = 2 * f x - f 0) (x y : ℤ) :
  f (x + y) = f x + f y - f 0 := by
  have h_main : f (x + y) = f x + f y - f 0 := by
    have h1 := h_f_all (x + y) 0
    have h2 := h_f_all x y
    have h3 := h_f_all (x + y) y
    have h4 := h_f_all x (x + y)
    have h5 := h_f_2x (x + y)
    have h6 := h_f_2x x
    have h7 := h_f_2x y
    have h8 := h_f_all 0 (x + y)
    have h9 := h_f_all 0 x
    have h10 := h_f_all 0 y
    have h11 := h_f_f_x (x + y)
    have h12 := h_f_f_x x
    have h13 := h_f_f_x y
    have h14 := h_f_all (2 * (x + y)) 0
    have h15 := h_f_all (2 * x) 0
    have h16 := h_f_all (2 * y) 0
    have h17 := h_f_all x 0
    have h18 := h_f_all y 0
    have h19 := h_f_all (x + y) (x + y)
    have h20 := h_f_all x x
    have h21 := h_f_all y y
    -- Simplify the expressions using the given conditions
    simp [h_f_2x, mul_add, add_mul, mul_comm, mul_left_comm, mul_assoc] at h1 h2 h3 h4 h5 h6 h7 h8 h9 h10 h11 h12 h13 h14 h15 h16 h17 h18 h19 h20 h21 ⊢
    <;> ring_nf at h1 h2 h3 h4 h5 h6 h7 h8 h9 h10 h11 h12 h13 h14 h15 h16 h17 h18 h19 h20 h21 ⊢
    <;> omega
  exact h_main

theorem prop_f_f_x (f : ℤ → ℤ) (h_f_all : ∀ a b, f (2 * a) + 2 * (f b) = f (f (a + b))) (x : ℤ) :
  f (f x) = 2 * f x + f 0 := by
  have h_main : f (f x) = 2 * f x + f 0 := by
    have h1 := h_f_all x 0
    have h2 := h_f_all 0 x
    have h3 := h_f_all x x
    have h4 := h_f_all (-x) x
    have h5 := h_f_all x (-x)
    have h6 := h_f_all 0 0
    have h7 := h_f_all x (-2 * x)
    have h8 := h_f_all (-x) (-x)
    have h9 := h_f_all x 1
    have h10 := h_f_all x (-1)
    have h11 := h_f_all 1 x
    have h12 := h_f_all (-1) x
    have h13 := h_f_all 1 0
    have h14 := h_f_all (-1) 0
    have h15 := h_f_all 0 1
    have h16 := h_f_all 0 (-1)
    have h17 := h_f_all 1 1
    have h18 := h_f_all (-1) (-1)
    -- Simplify the equations to find a relationship between f(0) and f(f(0))
    simp at h1 h2 h3 h4 h5 h6 h7 h8 h9 h10 h11 h12 h13 h14 h15 h16 h17 h18
    ring_nf at h1 h2 h3 h4 h5 h6 h7 h8 h9 h10 h11 h12 h13 h14 h15 h16 h17 h18 ⊢
    -- Use linear arithmetic to solve for the desired result
    omega
  exact h_main

theorem prop_f_2x (f : ℤ → ℤ) (h_f_all : ∀ a b, f (2 * a) + 2 * (f b) = f (f (a + b)))
    (h_f_f_x : ∀ x, f (f x) = 2 * f x + f 0) (x : ℤ) :
  f (2 * x) = 2 * f x - f 0 := by
  have h1 : f (f (2 * x)) = 2 * f (2 * x) + f 0 := by
    have h1 := h_f_f_x (2 * x)
    -- Simplify the expression using the given condition h_f_f_x
    simp at h1 ⊢
    <;> linarith
  
  have h2 : f (2 * x) + 2 * f x = f (f (2 * x)) := by
    have h2 := h_f_all x x
    -- Simplify the expression using the given condition h_f_all
    ring_nf at h2 ⊢
    <;> linarith
  
  have h3 : f (2 * x) + 2 * f x = 2 * f (2 * x) + f 0 := by
    have h3 : f (2 * x) + 2 * f x = f (f (2 * x)) := h2
    rw [h3]
    have h4 : f (f (2 * x)) = 2 * f (2 * x) + f 0 := h1
    rw [h4]
    <;> ring
    <;> omega
  
  have h4 : f (2 * x) = 2 * f x - f 0 := by
    have h5 : f (2 * x) + 2 * f x = 2 * f (2 * x) + f 0 := h3
    -- Rearrange the equation to isolate f(2 * x)
    have h6 : f (2 * x) = 2 * f x - f 0 := by
      -- Solve for f(2 * x) using linear arithmetic
      linarith
    exact h6
  
  apply h4


theorem cauchy_implies_linear_form (f : ℤ → ℤ) (h_cauchy_like : ∀ x y, f (x + y) = f x + f y - f 0) :
  ∃ c : ℤ, ∀ x, f x = c * x + f 0 := by
  have h_main : ∃ (c : ℤ), ∀ (x : ℤ), f x = c * x + f 0 := by
    use f 1 - f 0
    intro x
    have h1 : ∀ n : ℤ, f n = (f 1 - f 0) * n + f 0 := by
      intro n
      induction n using Int.induction_on with
      | hz =>
        -- Base case: n = 0
        simp [h_cauchy_like]
        <;> ring_nf
        <;> omega
      | hp n ih =>
        -- Inductive step: n = p + 1
        have h2 := h_cauchy_like n 1
        have h3 := h_cauchy_like 0 (n + 1)
        have h4 := h_cauchy_like (n + 1) 0
        have h5 := h_cauchy_like 1 0
        have h6 := h_cauchy_like 0 1
        simp at h2 h3 h4 h5 h6
        simp [ih, add_mul, mul_add, mul_one, mul_neg, mul_zero, sub_eq_add_neg] at h2 h3 h4 h5 h6 ⊢
        <;> ring_nf at *
        <;> omega
      | hn n ih =>
        -- Inductive step: n = - (n + 1)
        have h2 := h_cauchy_like (-n - 1) 1
        have h3 := h_cauchy_like 0 (-n - 1)
        have h4 := h_cauchy_like (-n - 1) 0
        have h5 := h_cauchy_like 1 0
        have h6 := h_cauchy_like 0 1
        simp at h2 h3 h4 h5 h6
        simp [ih, add_mul, mul_add, mul_one, mul_neg, mul_zero, sub_eq_add_neg] at h2 h3 h4 h5 h6 ⊢
        <;> ring_nf at *
        <;> omega
    have h2 := h1 x
    have h3 := h1 1
    have h4 := h1 0
    simp at h2 h3 h4 ⊢
    <;> linarith
  exact h_main

theorem step6_zero_function_is_solution (f : ℤ → ℤ) (h_zero : ∀ x, f x = 0) : (∀ a b, f (2 * a) + 2 * (f b) = f (f (a + b))) := by
have h_main : ∀ a b, f (2 * a) + 2 * (f b) = f (f (a + b)) := by
  intro a b
  have h1 : f (2 * a) = 0 := by
    rw [h_zero]
    <;> simp [h_zero]
  have h2 : f b = 0 := by
    rw [h_zero]
    <;> simp [h_zero]
  have h3 : f (a + b) = 0 := by
    rw [h_zero]
    <;> simp [h_zero]
  have h4 : f (f (a + b)) = 0 := by
    rw [h_zero]
    <;> simp [h_zero]
  -- Simplify the LHS and RHS using the above equalities
  simp [h1, h2, h3, h4, h_zero]
  <;> linarith
exact h_main

theorem step7_linear_function_is_solution (f : ℤ → ℤ) (c : ℤ) (h_lin : ∀ x, f x = 2 * x + c) : (∀ a b, f (2 * a) + 2 * (f b) = f (f (a + b))) := by
have h_main : ∀ a b, f (2 * a) + 2 * (f b) = f (f (a + b)) := by
  intro a b
  have h1 : f (2 * a) = 2 * (2 * a) + c := by
    rw [h_lin]
    <;> ring
  have h2 : f b = 2 * b + c := by
    rw [h_lin]
    <;> ring
  have h3 : f (f (a + b)) = f (2 * (a + b) + c) := by
    have h4 : f (a + b) = 2 * (a + b) + c := by
      rw [h_lin]
      <;> ring
    rw [h4]
    <;> ring
  have h4 : f (f (a + b)) = 2 * (2 * (a + b) + c) + c := by
    rw [h3]
    rw [h_lin]
    <;> ring
  have h5 : f (2 * a) + 2 * (f b) = (2 * (2 * a) + c) + 2 * (2 * b + c) := by
    rw [h1, h2]
    <;> ring
  have h6 : f (2 * a) + 2 * (f b) = 4 * a + 4 * b + 3 * c := by
    linarith
  have h7 : f (f (a + b)) = 4 * a + 4 * b + 3 * c := by
    linarith
  linarith
exact h_main

theorem imo2019_p1
    (f : ℤ → ℤ) :
    (∀ a b : ℤ, f (2 * a) + 2 * f b = f (f (a + b))) ↔ solution_set f := by
  constructor
  · intro h_fe
    have h_ff : ∀ x, f (f x) = 2 * f x + f 0 :=
      prop_f_f_x f h_fe
    have h_f2 : ∀ x, f (2 * x) = 2 * f x - f 0 :=
      prop_f_2x f h_fe h_ff
    have h_add : ∀ x y, f (x + y) = f x + f y - f 0 :=
      prop_cauchy_like f h_fe h_ff h_f2
    rcases cauchy_implies_linear_form f h_add with ⟨c, h_lin0⟩
    have h_split :
        (∀ x, f x = 0) ∨ (∀ x, f x = 2 * x + f 0) :=
      linear_form_plus_f_f_x_implies_solutions f c h_lin0 h_ff
    cases h_split with
    | inl h0 =>
        exact Or.inl h0
    | inr h2 =>
        exact Or.inr ⟨f 0, h2⟩
  · intro h_sol
    cases h_sol with
    | inl h0 =>
        exact step6_zero_function_is_solution f h0
    | inr h_exists =>
        rcases h_exists with ⟨c, h_lin⟩
        exact step7_linear_function_is_solution f c h_lin
\end{minted}

\subsection{IMO 2011 P3}
\begin{minted}[fontsize=\small, breaklines, linenos, bgcolor=gray!10]{lean}
-- Solution to IMO 2011 P3 by DRP-IMO

import Mathlib
import Aesop

set_option maxHeartbeats 0

open BigOperators Real Nat Topology Rat 



theorem imo2011_p3_lemma1_f_neg_le_self (f : ℝ → ℝ) (hf : ∀ x y, f (x + y) ≤ y * f x + f (f x)) :
  ∀ x, f x < 0 → f x ≤ x := by
  have h_main : ∀ (x : ℝ), f x < 0 → f x ≤ x := by
    intro x hx
    have h1 : f x ^ 2 - x * f x ≥ 0 := by
      have h2 := hf x (f x - x)
      have h3 := hf (f x) (x - f x)
      have h4 := hf x 0
      have h5 := hf 0 x
      have h6 := hf x x
      have h7 := hf x (-x)
      have h8 := hf (-x) x
      have h9 := hf 0 0
      have h10 := hf x 1
      have h11 := hf 1 x
      have h12 := hf x (-1)
      have h13 := hf (-1) x
      have h14 := hf x (f x)
      have h15 := hf (f x) x
      have h16 := hf x (-f x)
      have h17 := hf (-f x) x
      have h18 := hf x (x + f x)
      have h19 := hf (x + f x) x
      have h20 := hf x (-x)
      have h21 := hf (-x) x
      have h22 := hf x (x - f x)
      have h23 := hf (x - f x) x
      have h24 := hf x (f x + x)
      have h25 := hf (f x + x) x
      have h26 := hf x (2 * f x)
      have h27 := hf (2 * f x) x
      have h28 := hf x (-2 * f x)
      have h29 := hf (-2 * f x) x
      -- Normalize the expressions to simplify the inequalities
      ring_nf at h2 h3 h4 h5 h6 h7 h8 h9 h10 h11 h12 h13 h14 h15 h16 h17 h18 h19 h20 h21 h22 h23 h24 h25 h26 h27 h28 h29 ⊢
      -- Use linear arithmetic to prove the inequality
      nlinarith [sq_nonneg (f x - x), sq_nonneg (f x + x), sq_nonneg (f x - 2 * x), sq_nonneg (f x + 2 * x),
        sq_nonneg (2 * f x - x), sq_nonneg (2 * f x + x)]
    have h3 : f x ≤ x := by
      by_contra h
      have h4 : f x > x := by linarith
      have h5 : f x ^ 2 - x * f x < 0 := by
        nlinarith [hx, h4]
      nlinarith
    exact h3
  exact h_main

/--Let \( f : \mathbb{R} \to \mathbb{R} \) be a function such that for all real numbers \( x \) and \( y \), the inequality \( f(x + y) \leq y \cdot f(
x) + f(f(x)) \) holds. Prove that for all real numbers \( x \), the inequality \( f(x) \leq f(f(x)) \) is true.-/
theorem imo2011_p3_st1 (f : ℝ → ℝ) (hf : ∀ x y, f (x + y) ≤ y * f x + f (f x)) :
  ∀ x, f x ≤ f (f x) := by
  have h_main : ∀ (x : ℝ), f x ≤ f (f x) := by
    intro x
    have h₁ := hf x 0
    -- Simplify the inequality by substituting y = 0
    simp at h₁
    -- Use the simplified inequality to conclude the proof
    linarith
  exact h_main



/--Consider a function \( f : \mathbb{R} \to \mathbb{R} \) that satisfies the condition: for all real numbers \( x \) and \( y \), \( f(x + y) \leq y \
cdot f(x) + f(f(x)) \). Prove that for all real numbers \( x \), \( f(x) \leq 0 \).-/
theorem aux_f_nonpositive (f : ℝ → ℝ) (hf : ∀ x y, f (x + y) ≤ y * f x + f (f x)) : ∀ x, f x ≤ 0 := by
have h_main : ∀ x, f x ≤ 0 := by
  intro x
  by_contra h
  have h₁ : f x > 0 := by linarith
  have h₂ := hf x (-x)
  have h₃ := hf 0 (f x)
  have h₄ := hf x 0
  have h₅ := hf (-x) x
  have h₆ := hf (-x) (-x)
  have h₇ := hf x (f x)
  have h₈ := hf (f x) (-f x)
  have h₉ := hf (f x) 0
  have h₁₀ := hf 0 (-f x)
  have h₁₁ := hf x (2 * x)
  have h₁₂ := hf x (-2 * x)
  have h₁₃ := hf (2 * x) (-x)
  have h₁₄ := hf (2 * x) x
  have h₁₅ := hf (-2 * x) x
  have h₁₆ := hf (-2 * x) (-x)
  have h₁₇ := hf (f x) x
  have h₁₈ := hf (f x) (-x)
  have h₁₉ := hf x (f x)
  have h₂₀ := hf (-x) (f x)
  have h₂₁ := hf x (-f x)
  have h₂₂ := hf (-x) (-f x)
  have h₂₃ := hf (2 * f x) (-f x)
  have h₂₄ := hf (-2 * f x) (-f x)
  have h₂₅ := hf (2 * f x) (f x)
  have h₂₆ := hf (-2 * f x) (f x)
  have h₂₇ := hf (f x) (2 * f x)
  have h₂₈ := hf (f x) (-2 * f x)
  have h₂₉ := hf x (x)
  have h₃₀ := hf x (-x)
  have h₃₁ := hf 0 (2 * f x)
  have h₃₂ := hf 0 (-2 * f x)
  have h₃₃ := hf (2 * f x) 0
  have h₃₄ := hf (-2 * f x) 0
  have h₃₅ := hf (f x) (f x)
  have h₃₆ := hf (-f x) (f x)
  have h₃₇ := hf (f x) (-f x)
  have h₃₈ := hf (-f x) (-f x)
  norm_num at *
  <;>
  (try nlinarith) <;>
  (try linarith) <;>
  (try nlinarith [h₁, h₂, h₃, h₄, h₅, h₆, h₇, h₈, h₉, h₁₀, h₁₁, h₁₂, h₁₃, h₁₄, h₁₅, h₁₆, h₁₇, h₁₈, h₁₉, h₂₀, h₂₁, h₂₂, h₂₃, h₂₄, h₂₅, h₂₆, h₂₇, h₂₈, h₂
₉, h₃₀, h₃₁, h₃₂, h₃₃, h₃₄, h₃₅, h₃₆, h₃₇, h₃₈]) <;>
  (try
    nlinarith [hf 0 0, hf x 0, hf 0 x, hf x (-x), hf (-x) x, hf (x + x) (-x), hf (-x) (x + x), hf (x - x) (x + x), hf (x + x) (x - x)]) <;>
  (try
    nlinarith [hf 0 0, hf x 0, hf 0 x, hf x (-x), hf (-x) x, hf (x + x) (-x), hf (-x) (x + x), hf (x - x) (x + x), hf (x + x) (x - x)]) <;>
  (try
    nlinarith [hf 0 0, hf x 0, hf 0 x, hf x (-x), hf (-x) x, hf (x + x) (-x), hf (-x) (x + x), hf (x - x) (x + x), hf (x + x) (x - x)])
  <;>
  nlinarith

exact h_main



theorem lemma_final_implication (f : ℝ → ℝ) (hf : ∀ x y, f (x + y) ≤ y * f x + f (f x))
  (h_f_at_0_is_0 : f 0 = 0) (h_f_non_positive : ∀ x, f x ≤ 0) : ∀ x ≤ 0, f x = 0 := by
  have h_main : ∀ (x : ℝ), x ≤ 0 → f x = 0 := by
    intro x hx
    have h1 : f x = 0 := by
      by_cases hx0 : x = 0
      · -- If x = 0, then f(0) = 0 by hypothesis
        simp [hx0, h_f_at_0_is_0]
      · -- If x ≠ 0, then x < 0
        have hx1 : x < 0 := by
          cases' lt_or_gt_of_ne hx0 with h h
          · linarith
          · exfalso
            linarith
        -- Use the given inequality with y = x and y = -x to derive the desired result
        have h2 := hf x x
        have h3 := hf (-x) x
        have h4 := hf x (-x)
        have h5 := hf 0 x
        have h6 := hf x 0
        have h7 := hf 0 (-x)
        have h8 := hf (-x) 0
        -- Simplify the inequalities using the given conditions
        norm_num [h_f_at_0_is_0] at h2 h3 h4 h5 h6 h7 h8 ⊢
        nlinarith [h_f_non_positive x, h_f_non_positive (-x), h_f_non_positive (f x),
          h_f_non_positive (x + x), h_f_non_positive (x - x), h_f_non_positive 0]
    exact h1
  exact h_main

/--Let \( f : \mathbb{R} \to \mathbb{R} \) be a function satisfying the inequality \( f(x + y) \leq y \cdot f(x) + f(f(x)) \) for all real numbers \( x
 \) and \( y \). Suppose that \( f(0) = c \) and there exists some \( x_0 \) such that \( f(x_0) = 0 \). Prove that \( c \geq 0 \), \( f(c) = c \), \( 
f(y) \leq c \) for all real numbers \( y \), and \( f(y) \leq c \cdot y + c \) for all real numbers \( y \).-/
theorem lemma_properties_if_f_has_zero (f : ℝ → ℝ) (hf : ∀ x y, f (x + y) ≤ y * f x + f (f x))
  (h_f0_eq_c : f 0 = c) (hx₀ : ∃ x₀, f x₀ = 0) :
  c ≥ 0 ∧ f c = c ∧ (∀ y, f y ≤ c) ∧ (∀ y, f y ≤ c * y + c) := by
  have h_c_ge_zero : c ≥ 0 := by
    obtain ⟨x₀, hx₀⟩ := hx₀
    have h1 := hf x₀ (-x₀)
    have h2 := hf 0 (-x₀)
    have h3 := hf x₀ 0
    have h4 := hf 0 0
    have h5 := hf x₀ (f x₀)
    have h6 := hf 0 (f 0)
    have h7 := hf x₀ (-f x₀)
    have h8 := hf 0 (-f 0)
    norm_num [h_f0_eq_c, hx₀] at *
    <;>
    (try linarith) <;>
    (try nlinarith) <;>
    (try simp_all [h_f0_eq_c, hx₀]) <;>
    (try linarith) <;>
    (try nlinarith)
    <;>
    (try
      nlinarith [sq_nonneg (f x₀), sq_nonneg (f 0), sq_nonneg (x₀ + 0), sq_nonneg (x₀ - 0), sq_nonneg (f x₀ + f 0), sq_nonneg (f x₀ - f 0)])
    <;>
    (try
      nlinarith [sq_nonneg (f x₀), sq_nonneg (f 0), sq_nonneg (x₀ + 0), sq_nonneg (x₀ - 0), sq_nonneg (f x₀ + f 0), sq_nonneg (f x₀ - f 0)])
  
  have h_f_c_eq_c : f c = c := by
    obtain ⟨x₀, hx₀⟩ := hx₀
    have h₁ := hf x₀ (c - x₀)
    have h₂ := hf 0 (c)
    have h₃ := hf c (-c)
    have h₄ := hf x₀ 0
    have h₅ := hf 0 0
    have h₆ := hf x₀ (f x₀)
    have h₇ := hf 0 (f 0)
    have h₈ := hf x₀ (-x₀)
    have h₉ := hf 0 (-x₀)
    simp [h_f0_eq_c, hx₀] at h₁ h₂ h₃ h₄ h₅ h₆ h₇ h₈ h₉ ⊢
    <;>
    (try ring_nf at * <;> nlinarith) <;>
    (try
      {
        nlinarith [sq_nonneg (f x₀), sq_nonneg (c - x₀), sq_nonneg (c + x₀)]
      }) <;>
    (try
      {
        nlinarith [sq_nonneg (f x₀), sq_nonneg (c - x₀), sq_nonneg (c + x₀), sq_nonneg (f c)]
      })
    <;>
    (try
      {
        nlinarith [sq_nonneg (f x₀), sq_nonneg (c - x₀), sq_nonneg (c + x₀), sq_nonneg (f c), sq_nonneg (f 0)]
      })
    <;>
    (try
      {
        nlinarith [sq_nonneg (f x₀), sq_nonneg (c - x₀), sq_nonneg (c + x₀), sq_nonneg (f c), sq_nonneg (f 0), sq_nonneg (c - f x₀)]
      })
    <;>
    (try
      {
        nlinarith [sq_nonneg (f x₀), sq_nonneg (c - x₀), sq_nonneg (c + x₀), sq_nonneg (f c), sq_nonneg (f 0), sq_nonneg (c - f x₀), sq_nonneg (f x₀ - 
c)]
      })
  
  have h_f_le_c : ∀ y, f y ≤ c := by
    intro y
    have h1 := hf y (-y)
    have h2 := hf y (c - y)
    have h3 := hf 0 (y)
    have h4 := hf c (-c)
    have h5 := hf y 0
    have h6 := hf 0 0
    have h7 := hf c 0
    have h8 := hf 0 c
    have h9 := hf y (f y)
    have h10 := hf 0 (f 0)
    have h11 := hf y (-f y)
    have h12 := hf 0 (-f 0)
    have h13 := hf c (y - c)
    have h14 := hf 0 (y - c)
    have h15 := hf (y - c) c
    have h16 := hf (y - c) 0
    have h17 := hf (y - c) (f (y - c))
    have h18 := hf (y - c) (-f (y - c))
    norm_num [h_f0_eq_c, h_f_c_eq_c] at *
    <;>
    (try linarith) <;>
    (try nlinarith) <;>
    (try
      {
        nlinarith [sq_nonneg (f y - c), sq_nonneg (f 0), sq_nonneg (c), sq_nonneg (y), sq_nonneg (f c - c), sq_nonneg (f y)]
      }) <;>
    (try
      {
        nlinarith [sq_nonneg (f y - c), sq_nonneg (f 0), sq_nonneg (c), sq_nonneg (y), sq_nonneg (f c - c), sq_nonneg (f y), h_c_ge_zero]
      }) <;>
    (try
      {
        nlinarith [sq_nonneg (f y - c), sq_nonneg (f 0), sq_nonneg (c), sq_nonneg (y), sq_nonneg (f c - c), sq_nonneg (f y), h_c_ge_zero, sq_nonneg (f 
y)]
      }) <;>
    (try
      {
        nlinarith [sq_nonneg (f y - c), sq_nonneg (f 0), sq_nonneg (c), sq_nonneg (y), sq_nonneg (f c - c), sq_nonneg (f y), h_c_ge_zero, sq_nonneg (f 
y - c)]
      })
  
  have h_f_le_cy_add_c : ∀ y, f y ≤ c * y + c := by
    intro y
    have h₁ := h_f_le_c y
    have h₂ := h_f_le_c 0
    have h₃ := hf 0 y
    have h₄ := hf y 0
    have h₅ := hf y (-y)
    have h₆ := hf 0 (-y)
    have h₇ := hf y (c - y)
    have h₈ := hf 0 (c)
    have h₉ := hf c (-c)
    have h₁₀ := hf y 0
    have h₁₁ := hf 0 0
    have h₁₂ := hf y (f y)
    have h₁₃ := hf 0 (f 0)
    have h₁₄ := hf y (-f y)
    have h₁₅ := hf 0 (-f 0)
    have h₁₆ := hf c (y - c)
    have h₁₇ := hf 0 (y - c)
    have h₁₈ := hf (y - c) c
    have h₁₉ := hf (y - c) 0
    have h₂₀ := hf (y - c) (f (y - c))
    have h₂₁ := hf (y - c) (-f (y - c))
    norm_num [h_f0_eq_c, h_f_c_eq_c] at *
    <;>
    (try linarith) <;>
    (try nlinarith) <;>
    (try
      {
        nlinarith [h_c_ge_zero, sq_nonneg (f y - c), sq_nonneg (y), sq_nonneg (c - y), sq_nonneg (f y + c - c * y)]
      }) <;>
    (try
      {
        nlinarith [h_c_ge_zero, sq_nonneg (f y - c), sq_nonneg (y), sq_nonneg (c - y), sq_nonneg (f y + c - c * y), sq_nonneg (f y - c * y)]
      }) <;>
    (try
      {
        nlinarith [h_c_ge_zero, sq_nonneg (f y - c), sq_nonneg (y), sq_nonneg (c - y), sq_nonneg (f y + c - c * y), sq_nonneg (f y - c * y), sq_nonneg 
(f y - c)]
      })
    <;>
    (try
      {
        cases' le_total 0 y with hy hy <;>
        cases' le_total 0 (f y - c) with h h <;>
        cases' le_total 0 (c - y) with h' h' <;>
        nlinarith [h_c_ge_zero, sq_nonneg (f y - c), sq_nonneg (y), sq_nonneg (c - y), sq_nonneg (f y + c - c * y), sq_nonneg (f y - c * y)]
      })
    <;>
    nlinarith
  
  exact ⟨h_c_ge_zero, h_f_c_eq_c, h_f_le_c, h_f_le_cy_add_c⟩

theorem imo2011_p3 (f : ℝ → ℝ) (hf : ∀ x y, f (x + y) ≤ y * f x + f (f x)) : ∀ x ≤ 0, f x = 0 := by
  -- Step 1: Prove that f is non-positive everywhere.
  have h_nonpos : ∀ x, f x ≤ 0 := aux_f_nonpositive f hf

  -- Step 2: Prove that there must exist some x₀ such that f(x₀) = 0.
  -- We prove this by contradiction. Assume f(x) is never zero.
  have h_exists_zero : ∃ x₀, f x₀ = 0 := by
    by_contra h_no_zero
    -- The hypothesis from `by_contra` is `h_no_zero : ¬(∃ x₀, f x₀ = 0)`.
    -- We use `push_neg` to convert it into a more usable form.
    push_neg at h_no_zero
    -- Now, `h_no_zero : ∀ (x : ℝ), f x ≠ 0`.

    -- This, combined with `h_nonpos`, implies f(x) < 0 for all x.
    have h_always_neg : ∀ x, f x < 0 := fun x ↦ (h_nonpos x).lt_of_ne (h_no_zero x)

    -- From this, we can deduce f(0) = f(f(0)).
    have h_f0_eq_ff0 : f 0 = f (f 0) := by
      have h_ff0_neg : f (f 0) < 0 := h_always_neg (f 0)
      have hle : f (f 0) ≤ f 0 := imo2011_p3_lemma1_f_neg_le_self f hf (f 0) h_ff0_neg
      have hge : f 0 ≤ f (f 0) := imo2011_p3_st1 f hf 0
      linarith

    -- Now, we use the main inequality to derive a contradiction.
    -- Let x = f(0) and y = -f(0).
    specialize hf (f 0) (-f 0)

    -- Define a local lemma for `a + -a = 0` to ensure it's available.
    have add_neg_self_local : f 0 + -f 0 = 0 := by ring

    -- Rewrite the inequality step-by-step to derive the contradiction.
    rw [add_neg_self_local] at hf
    rw [← h_f0_eq_ff0] at hf
    rw [← h_f0_eq_ff0] at hf

    -- The inequality is now f 0 ≤ -(f 0)² + f 0, which implies 0 ≤ -(f 0)².
    have h_contr : 0 ≤ -(f 0) ^ 2 := by linarith [hf]

    -- This is a contradiction because f(0) < 0, so -(f 0)² < 0.
    have h_f0_neg : f 0 < 0 := h_always_neg 0
    have h_sq_pos : 0 < (f 0) ^ 2 := sq_pos_of_ne_zero (ne_of_lt h_f0_neg)
    linarith

  -- Step 3: Use the existence of a zero to prove f(0) = 0.
  obtain ⟨x₀, hx₀⟩ := h_exists_zero
  have h_f0_eq_0 : f 0 = 0 := by
    -- A lemma gives properties of f if it has a zero. One is f(0) ≥ 0.
    have h_props := lemma_properties_if_f_has_zero f hf rfl ⟨x₀, hx₀⟩
    have h_f0_nonneg : f 0 ≥ 0 := h_props.1
    -- Combining f(0) ≥ 0 with f(0) ≤ 0 (from h_nonpos) gives f(0) = 0.
    linarith [h_nonpos 0]

  -- Step 4: Now that we have f(x) ≤ 0 and f(0) = 0, apply the final lemma.
  exact lemma_final_implication f hf h_f0_eq_0 h_nonpos
\end{minted}

\subsection{IMO 2005 P3}
\begin{minted}[fontsize=\small, breaklines, linenos, bgcolor=gray!10]{lean}
-- Solution to IMO 2005 P3 by DRP-IMO

import Mathlib
import Aesop

set_option maxHeartbeats 0

open BigOperators Real Nat Topology Rat 


theorem inequality_part1_nonnegative (x y z : ℝ) (hx : x > 0) (hy : y > 0) (hz : z > 0) (h : x * y * z ≥ 1) :
  (x*x - 1/x + y*y - 1/y + z*z - 1/z) / (x*x + y*y + z*z) ≥ 0 := by
  have h_main : x*x + y*y + z*z - (1/x + 1/y + 1/z) ≥ 0 := by
    have h₁ : 0 < x * y := by positivity
    have h₂ : 0 < x * z := by positivity
    have h₃ : 0 < y * z := by positivity
    have h₄ : 0 < x * y * z := by positivity
    have h₅ : 0 < x * y * z * x := by positivity
    have h₆ : 0 < x * y * z * y := by positivity
    have h₇ : 0 < x * y * z * z := by positivity
    field_simp [hx.ne', hy.ne', hz.ne']
    rw [le_div_iff₀ (by positivity)]
    -- Use nlinarith to prove the inequality
    nlinarith [sq_nonneg (x - y), sq_nonneg (x - z), sq_nonneg (y - z),
      sq_nonneg (x * y - 1), sq_nonneg (x * z - 1), sq_nonneg (y * z - 1),
      mul_nonneg (sub_nonneg.mpr h) (sq_nonneg (x - y)),
      mul_nonneg (sub_nonneg.mpr h) (sq_nonneg (x - z)),
      mul_nonneg (sub_nonneg.mpr h) (sq_nonneg (y - z)),
      mul_nonneg (sub_nonneg.mpr h) (sq_nonneg (x * y - x * z)),
      mul_nonneg (sub_nonneg.mpr h) (sq_nonneg (x * y - y * z)),
      mul_nonneg (sub_nonneg.mpr h) (sq_nonneg (x * z - y * z))]
  
  have h_final : (x*x - 1/x + y*y - 1/y + z*z - 1/z) / (x*x + y*y + z*z) ≥ 0 := by
    have h₁ : x * x + y * y + z * z - (1 / x + 1 / y + 1 / z) ≥ 0 := h_main
    have h₂ : x * x + y * y + z * z > 0 := by positivity
    have h₃ : (x * x - 1 / x + y * y - 1 / y + z * z - 1 / z) / (x * x + y * y + z * z) ≥ 0 := by
      have h₄ : x * x - 1 / x + y * y - 1 / y + z * z - 1 / z = (x * x + y * y + z * z) - (1 / x + 1 / y + 1 / z) := by
        ring
      rw [h₄]
      have h₅ : ((x * x + y * y + z * z) - (1 / x + 1 / y + 1 / z)) / (x * x + y * y + z * z) ≥ 0 := by
        apply div_nonneg
        · linarith
        · linarith
      exact h₅
    exact h₃
  
  exact h_final


theorem inequality_part2_nonnegative (x y z : ℝ) (hx : x > 0) (hy : y > 0) (hz : z > 0) :
  ((x^5 - x^2)/(x^5 + y^2 + z^2) - (x*x - 1/x)/(x*x + y*y + z*z)) +
  ((y^5 - y^2)/(y^5 + z^2 + x^2) - (y*y - 1/y)/(y*y + z^2 + x*x)) +
  ((z^5 - z^2)/(z^5 + x^2 + y^2) - (z*z - 1/z)/(z*z + x*x + y*y)) ≥ 0 := by
  have h_main : ((x^5 - x^2)/(x^5 + y^2 + z^2) - (x*x - 1/x)/(x*x + y*y + z*z)) + ((y^5 - y^2)/(y^5 + z^2 + x^2) - (y*y - 1/y)/(y*y + z^2 + x*x)) + ((z
^5 - z^2)/(z^5 + x^2 + y^2) - (z*z - 1/z)/(z*z + x*x + y*y)) ≥ 0 := by
    have h₁ : (x^5 - x^2)/(x^5 + y^2 + z^2) - (x*x - 1/x)/(x*x + y*y + z*z) ≥ 0 := by
      have h₁₀ : 0 < x^5 + y^2 + z^2 := by positivity
      have h₁₁ : 0 < x*x + y*y + z*z := by positivity
      have h₁₂ : 0 < x^5 := by positivity
      have h₁₃ : 0 < x^3 := by positivity
      have h₁₄ : 0 < x^2 := by positivity
      have h₁₅ : 0 < x := by positivity
      field_simp
      rw [le_div_iff₀ (by positivity), ← sub_nonneg]
      ring_nf
      nlinarith [sq_nonneg (x^3 - x), sq_nonneg (x^2 - 1), sq_nonneg (x - 1),
        mul_nonneg hx.le (sq_nonneg (x^2 - 1)), mul_nonneg hx.le (sq_nonneg (x^3 - x)),
        mul_nonneg hx.le (sq_nonneg (x^2 - x)), mul_nonneg hx.le (sq_nonneg (x^3 - 1)),
        mul_nonneg (sq_nonneg (x - 1)) (sq_nonneg (x + 1)), mul_nonneg hx.le (sq_nonneg (x^2 - 2 * x + 1))]
    have h₂ : (y^5 - y^2)/(y^5 + z^2 + x^2) - (y*y - 1/y)/(y*y + z^2 + x*x) ≥ 0 := by
      have h₂₀ : 0 < y^5 + z^2 + x^2 := by positivity
      have h₂₁ : 0 < y*y + z^2 + x*x := by positivity
      have h₂₂ : 0 < y^5 := by positivity
      have h₂₃ : 0 < y^3 := by positivity
      have h₂₄ : 0 < y^2 := by positivity
      have h₂₅ : 0 < y := by positivity
      field_simp
      rw [le_div_iff₀ (by positivity), ← sub_nonneg]
      ring_nf
      nlinarith [sq_nonneg (y^3 - y), sq_nonneg (y^2 - 1), sq_nonneg (y - 1),
        mul_nonneg hy.le (sq_nonneg (y^2 - 1)), mul_nonneg hy.le (sq_nonneg (y^3 - y)),
        mul_nonneg hy.le (sq_nonneg (y^2 - y)), mul_nonneg hy.le (sq_nonneg (y^3 - 1)),
        mul_nonneg (sq_nonneg (y - 1)) (sq_nonneg (y + 1)), mul_nonneg hy.le (sq_nonneg (y^2 - 2 * y + 1))]
    have h₃ : (z^5 - z^2)/(z^5 + x^2 + y^2) - (z*z - 1/z)/(z*z + x*x + y*y) ≥ 0 := by
      have h₃₀ : 0 < z^5 + x^2 + y^2 := by positivity
      have h₃₁ : 0 < z*z + x*x + y*y := by positivity
      have h₃₂ : 0 < z^5 := by positivity
      have h₃₃ : 0 < z^3 := by positivity
      have h₃₄ : 0 < z^2 := by positivity
      have h₃₅ : 0 < z := by positivity
      field_simp
      rw [le_div_iff₀ (by positivity), ← sub_nonneg]
      ring_nf
      nlinarith [sq_nonneg (z^3 - z), sq_nonneg (z^2 - 1), sq_nonneg (z - 1),
        mul_nonneg hz.le (sq_nonneg (z^2 - 1)), mul_nonneg hz.le (sq_nonneg (z^3 - z)),
        mul_nonneg hz.le (sq_nonneg (z^2 - z)), mul_nonneg hz.le (sq_nonneg (z^3 - 1)),
        mul_nonneg (sq_nonneg (z - 1)) (sq_nonneg (z + 1)), mul_nonneg hz.le (sq_nonneg (z^2 - 2 * z + 1))]
    linarith
  exact h_main


-- The main theorem
theorem imo2005_p3 (x y z : ℝ) (hx : x > 0) (hy : y > 0) (hz : z > 0) (h_prod_ge_1 : x * y * z ≥ 1) :
  (x ^ 5 - x ^ 2) / (x ^ 5 + y ^ 2 + z ^ 2) + (y ^ 5 - y ^ 2) / (y ^ 5 + z ^ 2 + x ^ 2) + (z ^ 5 - z ^ 2) / (z ^ 5 + x ^ 2 + y ^ 2) ≥ 0 := by
  -- Define S_part1 (LHS of inequality_part1_nonnegative)
  let S_part1 := (x^2 - 1/x + y^2 - 1/y + z^2 - 1/z) / (x^2 + y^2 + z^2)

  -- Define S_part2 (LHS of inequality_part2_nonnegative)
  let S_part2 :=
    ((x^5 - x^2)/(x^5 + y^2 + z^2) - (x^2 - 1/x)/(x^2 + y^2 + z^2)) +
    ((y^5 - y^2)/(y^5 + z^2 + x^2) - (y^2 - 1/y)/(y^2 + z^2 + x^2)) +
    ((z^5 - z^2)/(z^5 + x^2 + y^2) - (z^2 - 1/z)/(z^2 + x^2 + y^2))

  -- Prove S_part1 ≥ 0 using inequality_part1_nonnegative
  have h_S_part1_nonneg : S_part1 ≥ 0 := by
    apply inequality_part1_nonnegative <;> assumption

  -- Prove S_part2 ≥ 0 using inequality_part2_nonnegative
  have h_S_part2_nonneg : S_part2 ≥ 0 := by
    apply inequality_part2_nonnegative <;> assumption

  -- Prove that the original LHS is equal to S_part1 + S_part2
  -- We'll prove it as a separate fact (have) and then use it.
  have h_LHS_eq_sum :
    (x^5 - x^2)/(x^5 + y^2 + z^2) + (y^5 - y^2)/(y^5 + z^2 + x^2) + (z^5 - z^2)/(z^5 + x^2 + y^2) =
    S_part2 + S_part1 := by
    -- Expand the definitions of S_part1 and S_part2
    unfold S_part1 S_part2
    -- Normalize denominators which are permutations of each other
    have h_denom_y : y^2 + z^2 + x^2 = x^2 + y^2 + z^2 := by ac_rfl
    have h_denom_z : z^2 + x^2 + y^2 = x^2 + y^2 + z^2 := by ac_rfl
    rw [h_denom_y, h_denom_z]
    -- The rest is a pure algebraic identity, which `ring` can solve.
    -- It correctly rearranges terms like (a-b)+(c-d)+(e-f) + (b+d+f)/k = a+c+e
    -- after combining the fractions for S_part1
    ring

  -- Rewrite the goal using the equality we just proved
  rw [h_LHS_eq_sum]

  -- The goal is now S_part2 + S_part1 ≥ 0, which follows from the two parts being non-negative.
  exact add_nonneg h_S_part2_nonneg h_S_part1_nonneg
\end{minted}

\subsection{IMO 2000 P2}
\begin{minted}[fontsize=\small, breaklines, linenos, bgcolor=gray!10]{lean}
-- Solution to IMO 2000 P2 by DRP-IMO

import Mathlib
import Aesop

set_option maxHeartbeats 0

open BigOperators Real Nat Topology Rat 

/--Given positive real numbers \( a \), \( b \), and \( c \) such that \( a \times b \times c = 1 \), prove that there exist positive real numbers \( x
 \), \( y \), and \( z \) such that \( a = \frac{x}{y} \), \( b = \frac{y}{z} \), and \( c = \frac{z}{x} \).-/
theorem imo2000_p2_existence_of_xyz (a b c : ℝ) (ha : 0 < a) (hb : 0 < b) (hc : 0 < c) (habc : a * b * c = 1) :
  ∃ x y z : ℝ, 0 < x ∧ 0 < y ∧ 0 < z ∧ a = x/y ∧ b = y/z ∧ c = z/x := by
  have h_main : ∃ (x y z : ℝ), 0 < x ∧ 0 < y ∧ 0 < z ∧ a = x/y ∧ b = y/z ∧ c = z/x := by
    refine' ⟨a, 1, 1 / b, _, _, _, _, _, _⟩
    · -- Prove that a > 0
      linarith
    · -- Prove that 1 > 0
      norm_num
    · -- Prove that 1 / b > 0
      exact div_pos zero_lt_one hb
    · -- Prove that a = a / 1
      field_simp
    · -- Prove that b = 1 / (1 / b)
      field_simp
      <;>
      nlinarith
    · -- Prove that c = (1 / b) / a
      have h₁ : c = 1 / (a * b) := by
        have h₂ : a * b * c = 1 := habc
        have h₃ : c = 1 / (a * b) := by
          have h₄ : a * b ≠ 0 := by positivity
          field_simp [h₄] at h₂ ⊢
          nlinarith
        exact h₃
      have h₂ : (1 / b : ℝ) / a = 1 / (a * b) := by
        field_simp
        <;> ring
        <;> field_simp [ha.ne', hb.ne']
        <;> nlinarith
      rw [h₁] at *
      <;> linarith
  exact h_main

/--Consider three positive real numbers \( x \), \( y \), and \( z \) such that \( x > 0 \), \( y > 0 \), and \( z > 0 \). Prove that the product of th
e expressions \( (x - y + z) \), \( (y - z + x) \), and \( (z - x + y) \) is less than or equal to the product \( x \cdot y \cdot z \).-/
theorem schur_like_ineq (x y z : ℝ) (hx : 0 < x) (hy : 0 < y) (hz : 0 < z) :
  (x - y + z) * (y - z + x) * (z - x + y) ≤ x * y * z := by
  have h_main : (x - y + z) * (y - z + x) * (z - x + y) ≤ x * y * z := by
    nlinarith [sq_nonneg (x - y), sq_nonneg (y - z), sq_nonneg (z - x),
      mul_nonneg hx.le hy.le, mul_nonneg hy.le hz.le, mul_nonneg hz.le hx.le,
      mul_nonneg (sq_nonneg (x - y)) (sq_nonneg (y - z)),
      mul_nonneg (sq_nonneg (y - z)) (sq_nonneg (z - x)),
      mul_nonneg (sq_nonneg (z - x)) (sq_nonneg (x - y)),
      mul_nonneg (sq_nonneg (x - y + z)) (sq_nonneg (y - z + x)),
      mul_nonneg (sq_nonneg (y - z + x)) (sq_nonneg (z - x + y)),
      mul_nonneg (sq_nonneg (z - x + y)) (sq_nonneg (x - y + z)),
      mul_nonneg (sq_nonneg (x + y - z)) (sq_nonneg (y + z - x)),
      mul_nonneg (sq_nonneg (y + z - x)) (sq_nonneg (z + x - y)),
      mul_nonneg (sq_nonneg (z + x - y)) (sq_nonneg (x + y - z))]
  exact h_main

/--Consider positive real numbers \( a, b, c, x, y, z \) such that \( a \cdot b \cdot c = 1 \) and \( a = \frac{x}{y} \), \( b = \frac{y}{z} \), \( c =
 \frac{z}{x} \). Prove that the inequality \((a - 1 + \frac{1}{b}) \cdot (b - 1 + \frac{1}{c}) \cdot (c - 1 + \frac{1}{a}) \leq 1\) is equivalent to th
e inequality \((x - y + z) \cdot (y - z + x) \cdot (z - x + y) \leq x \cdot y \cdot z\).-/
theorem inequality_equivalence_under_parametrization (a b c x y z : ℝ)
  (ha : 0 < a) (hb : 0 < b) (hc : 0 < c) (habc : a * b * c = 1)
  (hx : 0 < x) (hy : 0 < y) (hz : 0 < z)
  (hax : a = x / y) (hby : b = y / z) (hcz : c = z / x) :
  (a - 1 + 1 / b) * (b - 1 + 1 / c) * (c - 1 + 1 / a) ≤ 1 ↔
  (x - y + z) * (y - z + x) * (z - x + y) ≤ x * y * z := by
  have h_main : (a - 1 + 1 / b) * (b - 1 + 1 / c) * (c - 1 + 1 / a) = ((x + z - y) / y) * ((x + y - z) / z) * ((y + z - x) / x) := by
    have h₁ : a - 1 + 1 / b = (x + z - y) / y := by
      have h₁ : a = x / y := by linarith
      have h₂ : b = y / z := by linarith
      rw [h₁, h₂]
      field_simp [ha.ne', hb.ne', hx.ne', hy.ne', hz.ne']
      <;> ring_nf
      <;> field_simp [ha.ne', hb.ne', hx.ne', hy.ne', hz.ne']
      <;> nlinarith
    have h₂ : b - 1 + 1 / c = (x + y - z) / z := by
      have h₁ : b = y / z := by linarith
      have h₂ : c = z / x := by linarith
      rw [h₁, h₂]
      field_simp [ha.ne', hb.ne', hc.ne', hx.ne', hy.ne', hz.ne']
      <;> ring_nf
      <;> field_simp [ha.ne', hb.ne', hc.ne', hx.ne', hy.ne', hz.ne']
      <;> nlinarith
    have h₃ : c - 1 + 1 / a = (y + z - x) / x := by
      have h₁ : c = z / x := by linarith
      have h₂ : a = x / y := by linarith
      rw [h₁, h₂]
      field_simp [ha.ne', hb.ne', hc.ne', hx.ne', hy.ne', hz.ne']
      <;> ring_nf
      <;> field_simp [ha.ne', hb.ne', hc.ne', hx.ne', hy.ne', hz.ne']
      <;> nlinarith
    rw [h₁, h₂, h₃]
    <;> field_simp [ha.ne', hb.ne', hc.ne', hx.ne', hy.ne', hz.ne']
    <;> ring_nf
    <;> field_simp [ha.ne', hb.ne', hc.ne', hx.ne', hy.ne', hz.ne']
    <;> nlinarith
  
  have h_equiv : ((x + z - y) / y) * ((x + y - z) / z) * ((y + z - x) / x) ≤ 1 ↔ (x - y + z) * (y - z + x) * (z - x + y) ≤ x * y * z := by
    have h₁ : 0 < x * y := by positivity
    have h₂ : 0 < y * z := by positivity
    have h₃ : 0 < z * x := by positivity
    have h₄ : 0 < x * y * z := by positivity
    constructor
    · intro h
      have h₅ : ((x + z - y) / y) * ((x + y - z) / z) * ((y + z - x) / x) ≤ 1 := by linarith
      have h₆ : (x - y + z) * (y - z + x) * (z - x + y) ≤ x * y * z := by
        field_simp at h₅
        rw [div_le_one (by positivity)] at h₅
        nlinarith [sq_nonneg (x - y), sq_nonneg (y - z), sq_nonneg (z - x),
          mul_nonneg hx.le hy.le, mul_nonneg hy.le hz.le, mul_nonneg hz.le hx.le,
          mul_nonneg (sq_nonneg (x - y)) hz.le, mul_nonneg (sq_nonneg (y - z)) hx.le,
          mul_nonneg (sq_nonneg (z - x)) hy.le]
      linarith
    · intro h
      have h₅ : (x - y + z) * (y - z + x) * (z - x + y) ≤ x * y * z := by linarith
      have h₆ : ((x + z - y) / y) * ((x + y - z) / z) * ((y + z - x) / x) ≤ 1 := by
        field_simp
        rw [div_le_one (by positivity)]
        nlinarith [sq_nonneg (x - y), sq_nonneg (y - z), sq_nonneg (z - x),
          mul_nonneg hx.le hy.le, mul_nonneg hy.le hz.le, mul_nonneg hz.le hx.le,
          mul_nonneg (sq_nonneg (x - y)) hz.le, mul_nonneg (sq_nonneg (y - z)) hx.le,
          mul_nonneg (sq_nonneg (z - x)) hy.le]
      linarith
  
  have h_final : (a - 1 + 1 / b) * (b - 1 + 1 / c) * (c - 1 + 1 / a) ≤ 1 ↔ (x - y + z) * (y - z + x) * (z - x + y) ≤ x * y * z := by
    rw [h_main]
    rw [h_equiv]
    <;>
    simp_all
    <;>
    field_simp
    <;>
    ring_nf
    <;>
    nlinarith
  
  exact h_final

theorem imo2000_p2
    (a b c : ℝ) (ha : 0 < a) (hb : 0 < b) (hc : 0 < c)
    (habc : a * b * c = 1) :
    (a - 1 + 1 / b) * (b - 1 + 1 / c) * (c - 1 + 1 / a) ≤ 1 := by
  -- 1. Parametrize x y z using positive numbers
  obtain ⟨x, y, z, hx, hy, hz, ha_eq, hb_eq, hc_eq⟩ :=
    imo2000_p2_existence_of_xyz a b c ha hb hc habc
  -- 2. Use an equivalent lemma to transform the goal into the form involving x y z
  have h_equiv :=
    (inequality_equivalence_under_parametrization
        (a := a) (b := b) (c := c) (x := x) (y := y) (z := z)
        ha hb hc habc hx hy hz ha_eq hb_eq hc_eq)
  -- 3. The Schur-type inequality yields the conclusion on the right-hand side.
  have hxyz : (x - y + z) * (y - z + x) * (z - x + y) ≤ x * y * z :=
    schur_like_ineq x y z hx hy hz
  -- 4. Derive the original conclusion by reversing the equivalent proposition.
  exact h_equiv.mpr hxyz
\end{minted}

\end{document}